\documentclass[%
 reprint,
 amsmath,amssymb,
 aps,
prb,
floatfix,
]{revtex4-1}

\usepackage{ifpdf}
\usepackage{amsmath} 
\usepackage{amssymb}
\usepackage{amsfonts}
\usepackage{braket}
\usepackage{color}
\usepackage[colorlinks=true,linkcolor=blue,citecolor=blue,breaklinks]{hyperref}
\usepackage{graphicx}
\usepackage{dcolumn}
\usepackage{bm}

\newcommand{\la}{\langle}
\newcommand{\ra}{\rangle}

\newcommand{\im}{\text{Im}}
\newcommand{\re}{\text{Re}}

\newcommand{\NSR}{\text{NS/R}}

\usepackage{todonotes}

\begin{document}

\title{Out-of-time-ordered correlators in a quantum Ising chain}

\author{Cheng-Ju Lin}
\affiliation{Department of Physics and Institute for Quantum Information and Matter, California Institute of Technology, Pasadena, CA 91125, USA}
\author{Olexei I. Motrunich}
\affiliation{Department of Physics and Institute for Quantum Information and Matter, California Institute of Technology, Pasadena, CA 91125, USA}


\date{\today}

\begin{abstract}
Out-of-time-ordered correlators (OTOC) have been proposed to characterize quantum chaos in generic systems.
However, they can also show interesting behavior in integrable models, resembling the OTOC in chaotic systems in some aspects.
Here we study the OTOC for different operators in the exactly-solvable one-dimensional quantum Ising spin chain.
The OTOC for spin operators that are local in terms of the Jordan-Wigner fermions has a ``shell-like'' structure:
After the wavefront passes, the OTOC approaches its original value in the long-time limit, showing no signature of scrambling; the approach is described by a $t^{-1}$ power law at long time $t$.
On the other hand, the OTOC for spin operators that are nonlocal in the Jordan-Wigner fermions has a ``ball-like'' structure, with its value reaching zero in the long-time limit, looking like a signature of scrambling; the approach to zero, however, is described by a slow power law $t^{-1/4}$ for the Ising model at the critical coupling.
These long-time power-law behaviors in the lattice model are not captured by conformal field theory calculations.
The mixed OTOC with both local and nonlocal operators in the Jordan-Wigner fermions also has a ``ball-like'' structure, but the limiting values and the decay behavior appear to be nonuniversal.
In all cases, we are not able to define a parametrically large window around the wavefront to extract the Lyapunov exponent. 

\end{abstract}

\maketitle


\section{\label{sec:intro} Introduction}
First discussed by Larkin and Ovchinnikov\cite{a.i.larkin1969} and recently revived by Kitaev,\cite{alexeikitaev2015,kitaev2017} the out-of-time-ordered correlator (OTOC) has attracted a lot of attention in the physics community across many different fields, including quantum information, high-energy physics, and condensed matter physics.
Consider
\begin{eqnarray}
C_{WV}(t) &\equiv& \frac{1}{2} \langle [W(t), V]^\dagger [W(t), V] \rangle \nonumber \\
&=& \frac{1}{2} \left[ \langle V^\dagger W(t)^\dagger W(t) V \rangle + \langle W(t)^\dagger V^\dagger V W(t) \rangle \right. \nonumber \\
&&~~~ \left. - \langle W(t)^\dagger V^\dagger W(t) V \rangle - \langle V^\dagger W(t)^\dagger V W(t) \rangle \right] ~, \nonumber
\end{eqnarray}
where $\langle O \rangle \equiv \text{Tr}[e^{-\beta H} O] / \text{Tr}[e^{-\beta H}]$ denotes the thermal average and $W(t) \equiv e^{i H t} W e^{-i H t}$ is the Heisenberg evolution of the operator $W$.
We see that the last line involves operators with unusual time ordering, hence the name ``OTOC.''
In particular, if $W$ and $V$ are Hermitian and unitary (e.g.,\ Pauli matrices), then $C_{WV}(t) = 1 - \text{Re} F_{WV}(t)$, where $F_{WV}(t) \equiv \langle W(t) V W(t) V \rangle$.

There are several aspects about this object which make it interesting to study. 
First of all, such $C(t)$ is a possible diagnostic for quantum chaos.
In classical physics, one hallmark of chaos is that a small difference in the initial condition results in an exponential deviation of the trajectory---the famous ``butterfly effect.''
Denoting $q$ as the generalized coordinate of the classical system in the language of Hamiltonian dynamics, the butterfly effect can be diagnosed from the behavior $|\frac{\partial q(t)}{\partial q(0)}| \sim e^{\lambda_\text{L} t}$, where $\lambda_\text{L}$ is the Lyapunov exponent.
The object $\frac{\partial q(t)}{\partial q(0)}$ can be calculated from the Poisson bracket $\{q(t), p\}_\text{P.B.}$.\cite{maldacena2016,cotler2017}
A natural generalization of this diagnostic to quantum systems is by promoting the Poisson bracket to a commutator.
Therefore, the behavior of the object $C(t) = \langle |[x(t), p]|^2 \rangle \sim e^{2\lambda_\text{L} t}$ is an immediate generalization of the classical chaos to quantum systems, where using $|A|^2 \equiv A^\dagger A$ removes the effect of phase cancellations when averaging.
Unlike classical systems where $\lambda_L$ can be arbitrarily large, in quantum systems it was argued\cite{maldacena2016,scaffidi2017} that under some natural assumptions $\lambda_L$ is bounded by $2\pi/\beta$ (assuming the unit $\hbar=1$), where $\beta$ is the inverse temperature of the system.

Several works have used this diagnostic to argue for the existence of quantum butterfly effect\cite{aleiner2016,bohrdt2017,werman2017,shen2017} and extract the Lyapunov exponent, with examples including the $O(N)$ model,\cite{chowdhury2017} fermionic models with critical Fermi surface,\cite{Patel21022017} and weakly diffusive metals.\cite{patel2017}
On the other hand, some systems, for example Luttinger liquids\cite{dora2017} and many-body localized systems,\cite{huang2017a,swingle2017,chen2017a,fan2017,slagle2017} do not show the Lyapunov behavior and are hence characterized as less chaotic or as slow scramblers.
Also, some works have shown that in certain Hamiltonians, the exponent extracted from OTOC does not match the classical counterpart of the semiclassical limit.\cite{rozenbaum2017,hashimoto2017}
For systems with bounded local Hilbert space and Hamiltonians with local interactions, a work\cite{kukuljan2017a} proposed that the density-OTOC is a more suitable diagnostic. 

Another perspective on the OTOC is that it demonstrates the instability of the ``thermal field double state'' and the scrambling of information.\cite{shenker2014,roberts2015,maldacena2016}
It is expected that if $F(t)$ is small [or $C(t)$ is large] in the long-time limit, the system is scrambled; while large $F(t)$ [small $C(t)$] signals absence of scrambling.
This also leads to a more sophisticated quantum information-theoretical definition of scrambling.\cite{hosur2016}
There are also some considerations regarding the quasiprobability behind the OTOC.\cite{yunger-halpern2017,yunger_halpern_quasiprobability_2018}
Several works used holographic description to show the nontriviality of the OTOC.\cite{shenker2014,roberts2015a}
A conformal field theory calculation showed agreement with the holographic calculations.\cite{roberts2015}

From the operator point of view, $C(t)$ is a measure of operator spreading.
Let us consider a 1d quantum spin-1/2 chain for concreteness, and assume $W$ operates on site $i$ (denoted as $W_i$) while $V$ operates on site $j$ (denoted as $V_j$) which we will treat as a ``probe" and will vary its position.
The Heisenberg-evolved operator $W_i(t)$ can be written in the basis of Pauli-string operators, $W_i(t) = \sum_S a_S(t) S$, where $S$ runs over all Pauli-strings (e.g., $\dots \sigma_0^x \sigma_1^z\sigma_2^z \dots$) and $a_S(t)$ denotes the corresponding amplitudes.
Then, at infinite temperature, $C(t) = 2 \sum_S^\prime |a_S(t)|^2$, where the primed summation is over the Pauli-strings with nontrivial commutation with $V_j$, or $[S, V_j] \neq 0$, and for concreteness we assumed that such $[S, V_j]$ is a Pauli string itself (times 2), as is the case where the ``probing'' $V_j$ is a single-site Pauli operator.
Therefore, by examining $V_j$ at different positions, one can quantify to some degree how $W_i(t)$ is spread over the space.
Recent calculations in the case of the time evolution given by local random quantum gates show nontrivial operator spreading and OTOC growth.\cite{von_keyserlingk_operator_2018,nahum2017a,nahum_operator_2018,khemani2017,rakovszky2017,chapman2017}

While most of the works focus on the OTOC diagnosing scrambling in chaotic systems, it is also interesting to consider its behavior in nonchaotic or integrable systems.
From the operator spreading and information scrambling point of view, the OTOC in integrable systems could still be interesting and reveal some nontrivial aspects.
We therefore study in detail the OTOC in the quantum Ising chain
\begin{equation}
\label{eqn:IsingModel}
H = -\frac{J}{2} \left( \sum_{j=0}^{L-1} \sigma_j^z \sigma_{j+1}^z + g \sum_{j=0}^{L-1} \sigma_j^x \right) ~,
\end{equation}
with periodic boundary condition.
The specific choice of couplings is such that at the $T = 0$ quantum critical point, $g = 1$, the maximal quasiparticle velocity is $c = J$, and we will also set $J = 1$.
We will focus on the case where $W$ and $V$ are single-site Pauli matrices whose positions we can vary.
We will be interested in the quantities
\begin{equation}
C_{\mu\nu}(\ell, t) \equiv \frac{1}{2} \langle |[\sigma_\ell^\mu(t), \sigma_0^\nu]|^2 \rangle = 1 - \text{Re} F_{\mu\nu}(\ell, t) ~,
\end{equation}
where $\mu, \nu = x, y, z$, and $F_{\mu\nu}(\ell, t) = \langle \sigma_\ell^\mu(t) \sigma_0^\nu \sigma_\ell^\mu(t) \sigma_0^\nu \rangle$.
Using lattice translation and mirror (i.e., $j \to -j$) symmetries, one can easily show that $C_{\mu\nu}(\ell, t) = \frac{1}{2} \langle |[\sigma_0^\mu(t), \sigma_\ell^\nu]|^2 \rangle$.
In some occasions, it is more natural to consider the latter expression.

In particular, we will focus on $F_{xx}(\ell, t)$, $F_{zz}(\ell, t)$, and $F_{zx}(\ell, t)$, as they represent three different types of behavior of the OTOC in the quantum Ising chain.
The model is solved using Jordan-Wigner (JW) fermions.
In terms of these, some spin operators are local and some become nonlocal (i.e., contain string operator), and the three OTOCs correspond to different combinations of local and nonlocal operators.
Previous studies\cite{sachdev2011} have shown that there is a qualitative distinction between the dynamical correlation functions in the two cases.
For operators that are local in terms of the JW fermions, the correlations show power-law decay in time {\it at any temperature}.
On the other hand, correlations of nonlocal operators decay exponentially in time.
Thus, the nonlocal operators exhibit behavior that is closer to generic (i.e., nonintegrable) ``thermal'' behavior, in contrast to the local operators.
Similar distinction has also been observed in quench settings,\cite{rossini2010,calabrese2012,essler2016} where operators that are local in the JW fermions approach their limiting values in a power-law fashion (``slow thermalization''), while for the nonlocal operators the approach is exponential in time (``fast thermalization''); in both cases, the limiting values are described by a generalized Gibbs ensemble appropriate for this integrable model.
It is therefore interesting to see if such qualitatively different behavior has any nontrivial correspondence in the OTOC calculations.
Indeed, we observe that the OTOC composed with local operators shows no sign of scrambling, namely $\lim_{t \to \infty} F_{xx}(\ell, t) = 1$ (which is the same as the value at $t = 0$) and the approach is $t^{-1}$ power law.
On the other hand, the OTOC composed with nonlocal operators shows the signature of scrambling, $\lim_{t \to \infty} F_{zz}(\ell, t) \to 0$.
However, we find that the long-time behavior of $F_{zz}(\ell, t)$ is a very slow $t^{-1/4}$ power law; this is a departure from the exponential decays found in the dynamical correlation and quench settings described above and shows that the OTOC encodes some different aspects; the very slow decay is also highly unusual and not fully understood.

The paper is organized as follows.
In Sec.~\ref{sec:diagIsing}, we briefly state the procedure of diagonalizing the Hamiltonian and establish some basic notations.
In Secs.~\ref{sec:xxotoc}, \ref{sec:zzotoc}, and \ref{sec:zxotoc}, we present the results for $C_{xx}(\ell, t)$, $C_{zz}(\ell, t)$, and $C_{zx}(\ell, t)$ respectively (details of the calculations are in Appendices~\ref{app:Pf_Fxx},~\ref{app:Pf_Fzz}, and \ref{app:Pf_Fzx} respectively).
In each case, we discuss the behavior at short time (spacelike region), behavior around the wavefront, and behavior at long time (timelike region).
In Appendix~\ref{app:sigmax_evolution}, we provide additional intuition about the $C_{xx}(\ell, t)$ and $C_{zx}(\ell, t)$ directly from the operator spreading picture by extracting these from the $\sigma^x(t)$ operator.
Finally, in Sec.~\ref{sec:conclusion}, we summarize and discuss some outstanding questions and future directions.

\section{Diagonalizing the Hamiltonian and setting up OTOC calculations}\label{sec:diagIsing}
We consider the quantum Ising model, Eq.~(\ref{eqn:IsingModel}), on a finite chain with periodic boundary conditions used to minimize boundary effects.
We diagonalize the model via Jordan-Wigner transformation and subsequent Bogoliubov transformation.\cite{sachdev2011}
In the fermionic representation, the spin operators are written as 
$\sigma_j^x = 1 - 2 c_j^\dagger c_j$ and 
$\sigma_j^z = -\prod_{j'<j}(1 - 2 c_{j'}^\dagger c_{j'}) (c_j + c_j^\dagger)$. 
We therefore obtain
\begin{eqnarray}
H &=& H_\text{NS} P_+ + H_\text{R} P_- ~, \label{eqn:HintoNSandR} \\
H_{\text{NS}/\text{R}} &=& -\frac{J}{2} \sum_{j=0}^{L-1} \left( c_j^\dagger c_{j+1} + c_{j+1}^\dagger c_j + c_j^\dagger c_{j+1}^\dagger + c_{j+1} c_j \right. \nonumber \\
&& ~~~~~~~~~~~ \left. - 2 g c_j^\dagger c_j + g \right) ~,
\end{eqnarray}
where $P_\pm = [1 \pm (-1)^{N_\text{tot}}]/2$ are the projectors to even/odd fermion number parity sectors, with $N_\text{tot} = \sum_{j=0}^{L-1} c_j^\dagger c_j$ the total fermion number;
$H_\text{NS}$ is understood with $c_{j+L} = -c_j$ boundary conditions (Neveu-Schwarz boundary conditions), while in $H_\text{R}$ we have $c_{j+L} = c_j$ (Ramond boundary conditions).
We then use appropriate Fourier transform $c_k = \frac{1}{\sqrt{L}} \sum_j c_j e^{-i k j}$ for each Hamiltonian $H_{\text{NS}/\text{R}}$ and Bogoliubov transformation $\gamma_k = u_k c_k - i w_k c^\dagger_{-k}$, diagonalizing $H_{\text{NS}/\text{R}} = \sum_{k \in K_{\text{NS}/\text{R}}} \epsilon_k (\gamma_k^\dagger \gamma_k - \frac{1}{2})$, where 
$K_\text{NS} = \{\frac{(2n+1) \pi}{L} | n \!=\! 0, \dots, (L\!-\!1) \}$ and 
$K_\text{R} = \{\frac{2 n \pi}{L} | n \!=\! 0, \dots, (L\!-\!1) \}$;
the quasiparticle dispersion is $\epsilon_k = J (1 + g^2 - 2 g \cos k)^{1/2}$.
This diagonalization is achieved by choosing the coherence factors as $u_k = \cos(\theta_k/2)$ and $w_k = \sin(\theta_k/2)$, where $\tan(\theta_k) = \sin(k)/[g - \cos(k)]$.

When making connections with the spin model, particularly when dealing with the string operators, it will be convenient to use Majorana representation. 
We will follow Ref.~\onlinecite{sachdev2011} and introduce Majorana fermions $A_j \equiv c_j^\dagger + c_j$ and $B_j = c_j^\dagger - c_j$.

The OTOCs can thus be expressed as fermionic correlation functions.
The thermal ensemble grants the Wick's theorem, which allows us to express all the correlation functions using two-point correlation functions.
However, the different fermion boundary conditions in the different fermion-number-parity sectors results in some complications in the calculation of dynamical correlation functions, and a more sophisticated treatment is needed.
We will carefully state the procedure below and in the subsequent sections for the specific OTOCs.
To prepare for such discussion, we here introduce some notations which will be useful later.

To enable free-fermion calculations, we introduce thermal ensembles corresponding to the two types of boundary conditions, $Z_\NSR \equiv \text{Tr}(e^{-\beta H_\NSR})$ and $\langle O \rangle_\NSR \equiv \text{Tr}(e^{-\beta H_\NSR} O)/Z_\NSR$.
Note that the trace in each case is defined over the full Fock space, i.e., including both parity sectors, even though $H_\NSR$ originally arose in the even/odd parity sectors.
These ensembles are introduced because the Wick's theorem only holds for an ensemble defined with respect to a quadratic Hamiltonian that is fixed over the full Fock space.
To evaluate the thermal average with respect to the spin Hamiltonian $H$, we recall Eq.~(\ref{eqn:HintoNSandR}) and use
\begin{equation}\label{eqn:secProj}
\langle O \rangle = \frac{Z_\text{NS}}{Z} \langle O P_+ \rangle_\text{NS} + \frac{Z_\text{R}}{Z} \langle O P_- \rangle_\text{R} ~.
\end{equation}
Since $P_\pm = [1 \pm (-1)^{N_\text{tot}}]/2$, we will have to calculate $\langle O \rangle_{\text{NS}/\text{R}}$ and $\langle O (-1)^{N_\text{tot}} \rangle_{\text{NS}/\text{R}}$.

We are interested in situations where $O$ in Eq.~(\ref{eqn:secProj}) is composed of several time evolved operators, $O = Q_1(t_1) Q_2(t_2) \dots$, where $Q(t) = e^{i H t} Q e^{-i H t}$.
To be able to use free-fermion calculations and Wick's theorem, it is crucial to require that each $Q_1, Q_2, \dots,$ does not change the fermion parity.
In this case, considering, e.g., the operator in the first term in Eq.~(\ref{eqn:secProj}), we have: 
$O P_+ = Q_1(t_1) Q_2(t_2) \dots P_+ = Q_1^\text{NS}(t_1) Q_2^\text{NS}(t_2) \dots P_+$, where $Q^\text{NS}(t) \equiv e^{i H_\text{NS} t} Q e^{-i H_\text{NS} t}$.
At this point, we can evaluate 
$\langle O P_+ \rangle_\text{NS} = \langle Q_1^\text{NS}(t_1) Q_2^\text{NS}(t_2) \dots P_+ \rangle _\text{NS} = [\langle Q_1^\text{NS}(t_1) Q_2^\text{NS}(t_2) \dots \rangle_\text{NS} + \langle Q_1^\text{NS}(t_1) Q_2^\text{NS}(t_2) \dots (-1)^{N_\text{tot}} \rangle_\text{NS}]/2$.
For each term in the last expression, both the the density matrix and the time evolution are determined by $H_\text{NS}$ viewed over the full Fock space (i.e., including both parity sectors), thus enabling free-fermion calculations.
Similar considerations apply to the calculation of $\langle O P_- \rangle_\text{R}$, which can be expressed entirely in terms of free fermions with Hamiltonian $H_\text{R}$ over the full Fock space.
We will often abuse the notation by dropping the labels ``NS'' or ``R'' in $Q^\text{NS}(t)$ or $Q^\text{R}(t)$ for brevity where the precise meaning can be recovered from the context.

In the thermodynamic limit, one in fact expects $\langle O (-1)^{N_\text{tot}} \rangle_{\text{NS}/\text{R}} \rightarrow 0$ and $\langle O \rangle_\text{NS} = \langle O \rangle_\text{R} = \langle O \rangle$. 
While for all the calculations one can in principle just evaluate $\langle O \rangle_\text{NS}$ or $\langle O \rangle_\text{R}$ and take the thermodynamical limit, in this paper we calculate exact finite-size $F_{\mu\nu}(\ell, t)$ [and hence $C_{\mu\nu}(\ell, t)$] using Eq.~(\ref{eqn:secProj}) so that we can compare the results against exact diagonalization of the spin system at small system sizes to ensure the correctness. 

To study the behavior of $C_{\mu\nu}(\ell, t)$ around the wavefront in more detail, or more specifically, to examine if the wavefront has the functional form $C_{\mu\nu}(\ell, t) \sim e^{-\lambda (\ell - c t)}$, we will also study the function $G_{\mu\nu}(\ell, t) = \partial \ln C_{\mu\nu}(\ell, t)/\partial t$, which characterizes the onset of the scrambling\cite{maldacena2016} if there is one and the spreading of the operator wavefront\cite{von_keyserlingk_operator_2018, nahum2017a, khemani2017}.
When discussing the analytical results and the calculation of $G_{\mu\nu}(\ell, t)$, we consider only the part $\langle O \rangle_\text{NS}$.

The crucial ingredients to obtain all the correlation functions are the two-point Majorana correlation functions, which we list in Appendix~\ref{app:correlation}.
In all the calculations of the OTOC, we will need to use the numerical values of $Z$, $Z_\text{NS}$, and $Z_\text{R}$.
The partition sums $Z_\text{NS}$ and $Z_\text{R}$ can be calculated easily by $Z_\NSR = \sum_{E_\NSR} e^{-\beta E_\NSR}$, where $E_\NSR$ denotes the eigenenergies of $H_\NSR$.
Note again that here we consider $H_\text{NS}$ acting on the full fermion Fock space including both even and odd parity sectors and performs free-fermion calculation of $Z_\text{NS}$, and similarly treats $H_\text{R}$ to calculate $Z_\text{R}$.
On the other hand, the calculation of $Z$ is nontrivial as it involves the projectors to the different sectors, and its details are presented in Appendix~\ref{app:correlation}.

\section{XX OTOC}\label{sec:xxotoc}
First, we discuss the commutator function $C_{xx}(\ell, t) = 1 - \text{Re} F_{xx}(\ell, t)$.
In the fermionic representation, $\sigma^x_\ell = A_\ell B_\ell$. 
Therefore we have
\begin{equation}\label{eqn:Fxx_in_Majorana}
F_{xx}(\ell, t) = \langle A_\ell(t) B_\ell(t) A_0 B_0 A_\ell(t) B_\ell(t) A_0 B_0 \rangle ~.
\end{equation}
Note that we need to use Eq.~(\ref{eqn:secProj}) and evaluate both $\langle O P_+ \rangle_\text{NS}$ and $\langle O P_- \rangle_\text{R}$.
While these expectation values can be evaluated using Wick's theorem, the calculation is simplified when cast in the form of Pfaffians of anitsymmetric matrices.
We present details in Appendix~\ref{app:Pf_Fxx}.

Figure~\ref{fig:XXotoc_all} shows the numerical results for $C_{xx}(\ell, t)$ at various time slices.
We can immediately identify the velocity of the wavefront as $c = 1$, which is the maximum of the quasiparticle group velocity $v_k = \partial \epsilon_k /\partial k$.
In the present case, the OTOC function is ``shell-like.''
That is, inside the timelike region, in the long-time limit, $C_{xx}(\ell, t) \to 0$, indicating no scrambling.
More precisely, as far as characterizing the operator spreading of $\sigma^x(t)$, the vanishing of the $C_{xx}$ OTOC in the long-time limit suggests that expansion of $\sigma^x(t)$ in terms of Pauli strings does not contain many $\sigma^y$ or $\sigma^z$ operators ``in the middle'' of the strings.
This can be indeed seen from the explicit expressions for $\sigma^x(t)$ in Appendix~\ref{app:sigmax_evolution}.

\begin{figure}
\includegraphics[width=1\columnwidth]{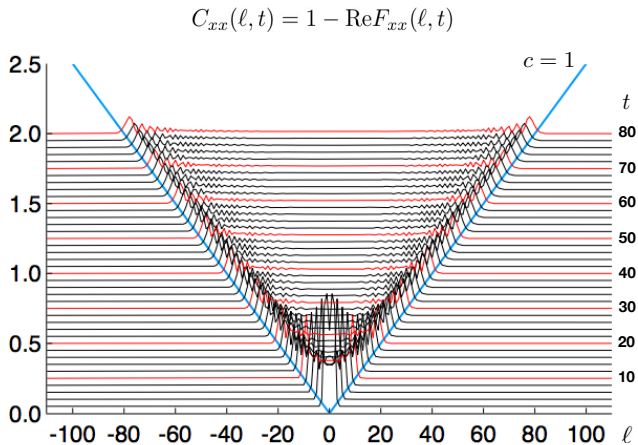}
\caption{\label{fig:XXotoc_all}
(color online) The function $C_{xx}(\ell, t)$ for the quantum Ising chain at the critical point, $g = 1$, at infinite temperature (inverse temperature $\beta = 0$); the system size is $L = 512$.
We show data as a function of $\ell$ at fixed time $t$, for $t$ in steps of $\Delta t = 2$ marked along the right border; here and in all figures, the energy unit $J$ in Eq.~(\ref{eqn:IsingModel}) is set to $1$.
The traces at fixed $t$ are shifted in the $y$ direction by $0.025 t$ thus offering three-dimensional-like visualization.
For every $t$ that is a multiple of $10$, we mark the trace with red color for easier reading of the data.
The light cone can be readily identified and corresponds to the maximal quasiparticle group velocity $c =
\max_k \frac{d \epsilon_k}{d k} = J = 1$.
In the timelike region, $C_{xx}(\ell, t)$ approaches zero in the long-time limit, indicating the absence of ``scrambling.''
}
\end{figure}

\subsection{``Universal'' early-time growth with separation-dependent power law}
\begin{figure}
\includegraphics[width=1\columnwidth]{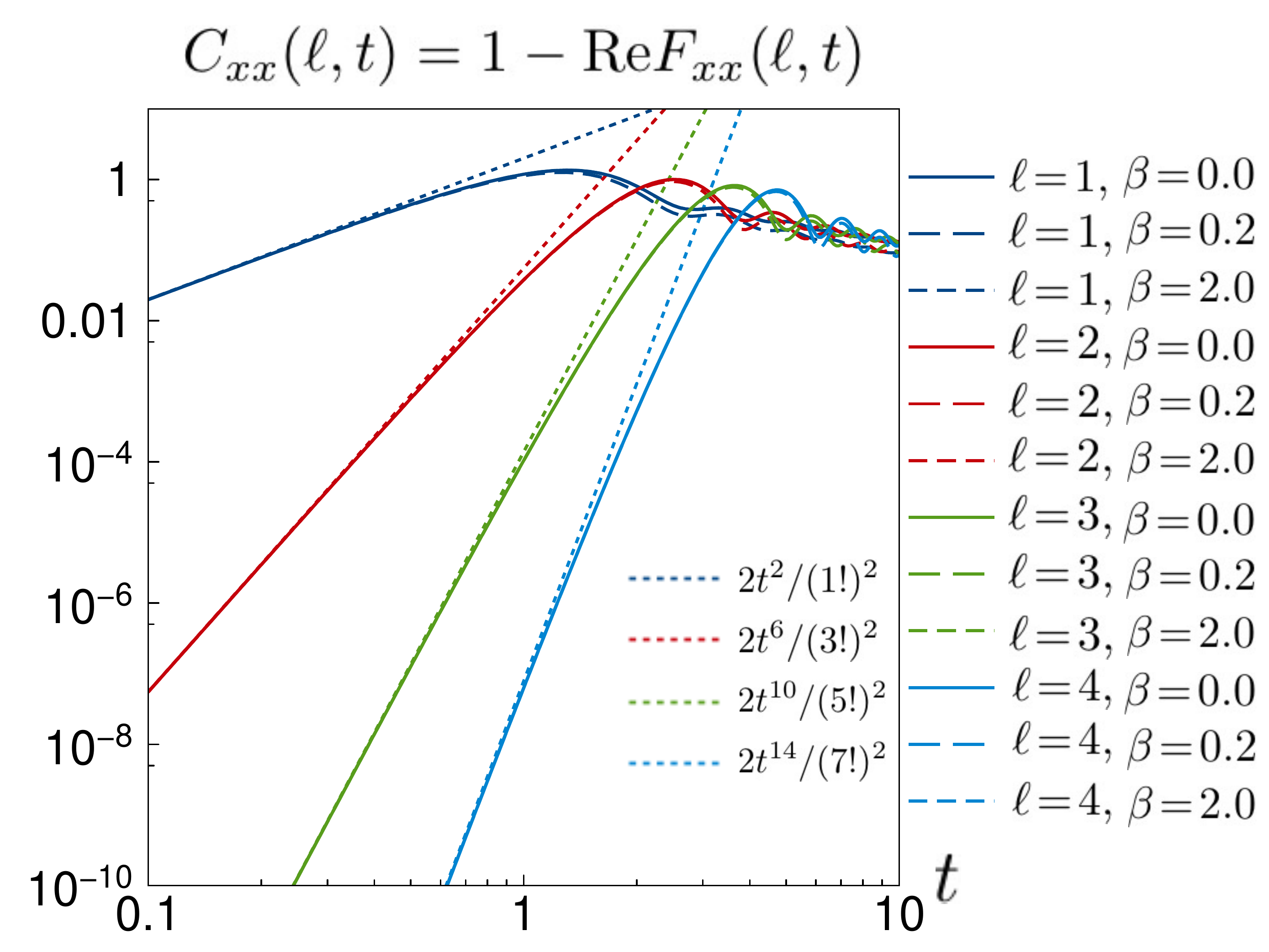}
\caption{\label{fig:xxotoc_short}
(color online) The function $C_{xx}(\ell, t)$ for several fixed separations $\ell$ at short time before the light cone reaches (i.e., spacelike separation between the operators).
The growth of the commutator is compared to the ``universal'' power-law behavior given by $\approx 2 t^{2 (2\ell - 1)}/[(2\ell \!-\! 1)!]^2$; note that there is essentially no temperature dependence in this regime.
}
\end{figure}

Before the light cone reaches, we can argue that there is a ``universal'' power-law growth of $C_{xx}(\ell, t) \sim t^{2 (2\ell - 1)}$.
Indeed, consider $W = \sigma^x_0$ and $V = \sigma^x_\ell$.
The Heisenberg evolution $W(t)$ at short time can be expanded via Hausdorff-Baker-Campbell (HBC) formula 
\begin{equation}\label{eqn:HBC}
W(t) = \sum_{n=0}^\infty \frac{t^n}{n!} L^n(W) ~,
\end{equation}
where $L(W) \equiv i[H, W]$. 
It is easy to check that for these $W$ and $V$, the smallest $n$ such that $[L^n(W), V] \neq 0$ is $n = 2\ell \! - \! 1$, and the nonzero contribution to the commutator comes from the piece in $L^n(W)$ that reaches site $\ell$, namely $J^{2\ell-1} g^{\ell-1} \sigma^y_0 \sigma^x_1 \dots \sigma^x_{\ell-1} \sigma^z_\ell$.
Therefore, the leading order behavior is
\begin{equation}
C_{xx}(\ell, t) \approx \frac{2 (Jt)^{4\ell-2} g^{2\ell-2}}{[(2\ell-1)!]^2} ~,
\end{equation}
which is also shown in Fig.~\ref{fig:xxotoc_short} and captures well the exact calculation in this regime.
We expect that such an argument based on the HBC formula is in fact very general and not related to any integrability of the model.\cite{roberts2016,dora2017,chen2017}
We thus expect such power-law growth with position-dependent power to be ``universal,'' present also in nonintegrable systems, as long as one is considering systems with bounded on-site Hilbert spaces and Hamiltonians with local interactions.
Such a power-law growth is indeed also observed in the XXZ model.\cite{dora2017}
However, we emphasize that this is just a quantum mechanical effect before the light cone reaches and should not be identified as a signature of scrambling or lack of it.

Lastly, we note that if we fix time $t$ and take the separation $\ell$ to large values, the commutator function $C_{xx}(\ell,t)$ decays faster than the exponential function in $\ell$, namely $C_{xx}(\ell,t)\sim \exp[a(t)\ell-4\ell \ln \ell]$, where $a(t)$ is some number that depends on $t$.

\subsection{Behavior around the wavefront}
\begin{figure}
\includegraphics[width=1\columnwidth]{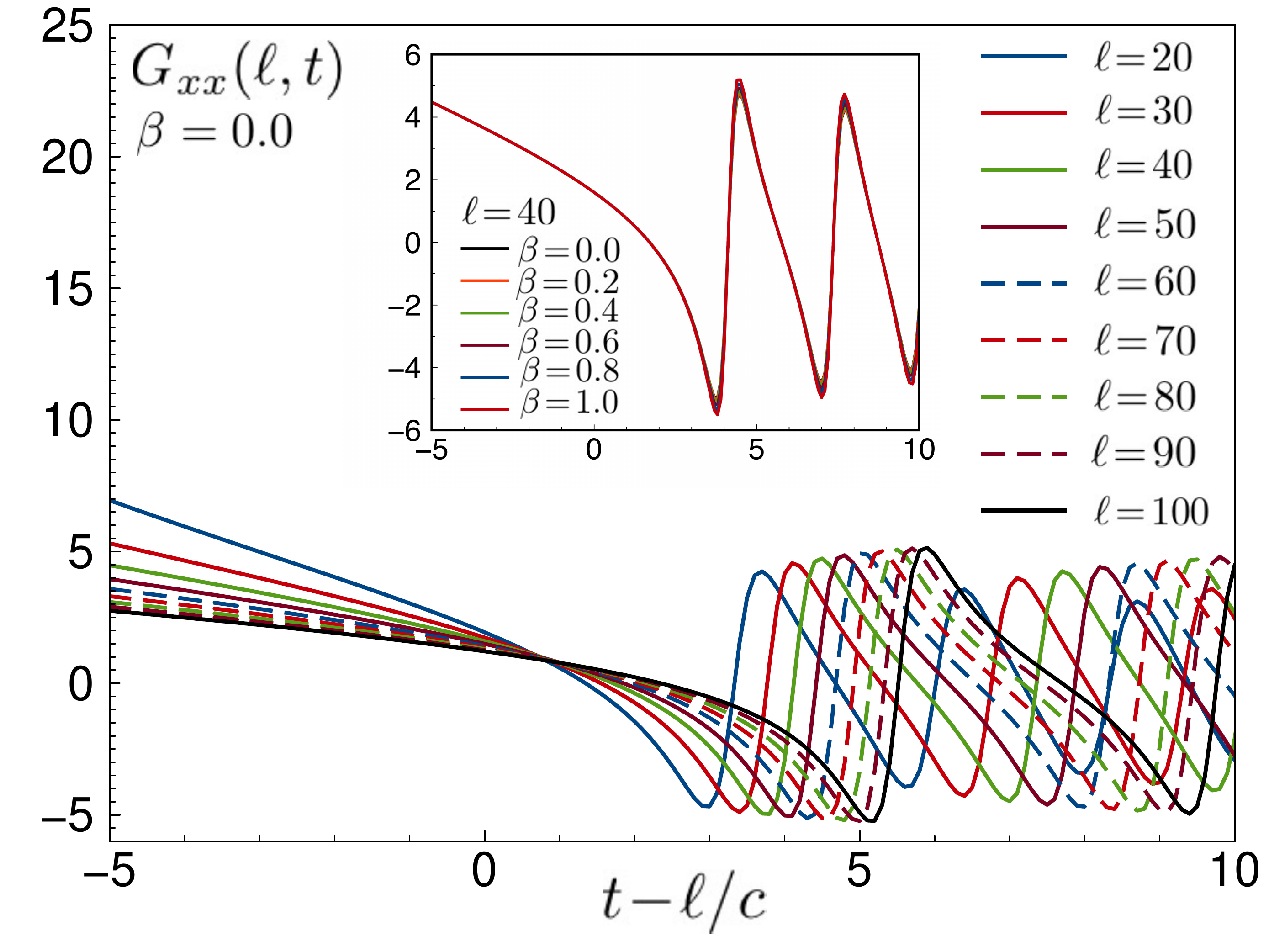}
\caption{\label{fig:xxotoc_wavefront}
The derivative function $G_{xx}(\ell, t) \equiv \partial \ln C_{xx}(\ell, t)/\partial t$ around the wavefront.
Before the oscillation sets in, $G_{xx}(\ell, t)$ has very strong $\ell$ dependence, for which we do not know any universal description.
Inset: $G_{xx}(\ell, t)$ for fixed $\ell = 40$ and several different inverse temperatures $\beta$, illustrating that there is basically no temperature dependence around the wavefront.
}
\end{figure}

To examine the behavior of $C_{xx}(\ell, t)$ around the wavefront more closely, we study the function 
\begin{equation}
G_{xx}(\ell, t) \equiv \frac{\partial \ln C_{xx}(\ell, t)}{\partial t} ~.
\end{equation}
We can calculate this in a way that avoids numerical differentiation (see Appendix~\ref{app:Pf_Fxx} for details) and present the results in Fig.~\ref{fig:xxotoc_wavefront}.
We see that before the oscillation sets in, $G_{xx}(\ell, t)$ shows very strong $\ell$ dependence.
On the other hand, the inset in Fig.~\ref{fig:xxotoc_wavefront} demonstrates that $G_{xx}(\ell, t)$ shows essentially no temperature dependence.
We conclude that the behavior near this wavefront does not show the ``exponential divergence'' that could be associated with the ``butterfly effect,'' and we can exclude the possibility of any temperature-dependent description of the wavefront.
Thinking about possible other descriptions of the wavefront, we do not clearly see a parametrically large time window where we could sharply distinguish this transition region behavior from the short-time and long-time behaviors.
While we see that the onset of oscillations (more precisely, onset of nonmonotonic behavior) happens at larger $t - \ell/c$ when $\ell$ is increased, at present we do not know if there is any asymptotic functional form in a well-defined window to describe the wavefront.
Thus we also note that the frequency of oscillations vanishes as one approaches the $\ell/t = c$ ray, so the later ``onset'' of oscillations for larger $\ell$ could be related to this.  
In any case, we can definitely tell that any ``universal'' description needs to be essentially temperature-independent.

\subsection{Universal long-time decay with $t^{-1}$ power law}
\label{subsec:Cxx_longtime}
The limiting value of $F_{xx}(\ell, t)$ for fixed $\ell$ but $t \to \infty$ can be easily shown to be one. 
Indeed, considering all the Wick contractions in Eq.~(\ref{eqn:Fxx_in_Majorana}), we see that if the contraction has any nonequal time correlation function, this term will be zero since all the fermionic correlation functions go to zero in the $t \to \infty$ limit.
We therefore have 
\begin{eqnarray}
F_{xx}(\ell, \infty) &=& \langle A_\ell(\infty) B_\ell(\infty) A_\ell(\infty) B_\ell(\infty) \rangle 
\langle A_0 B_0 A_0 B_0 \rangle \nonumber \\
&=& \left(\langle A_0 B_0 \rangle^2 + 1 + \langle A_0 B_0 \rangle \langle B_0 A_0 \rangle \right)^2 = 1 ~. \nonumber
\end{eqnarray}
We conclude that $C_{xx}(\ell, \infty) = 0$, which is a signature of no scrambling. 

\begin{figure}
\includegraphics[width=1\columnwidth]{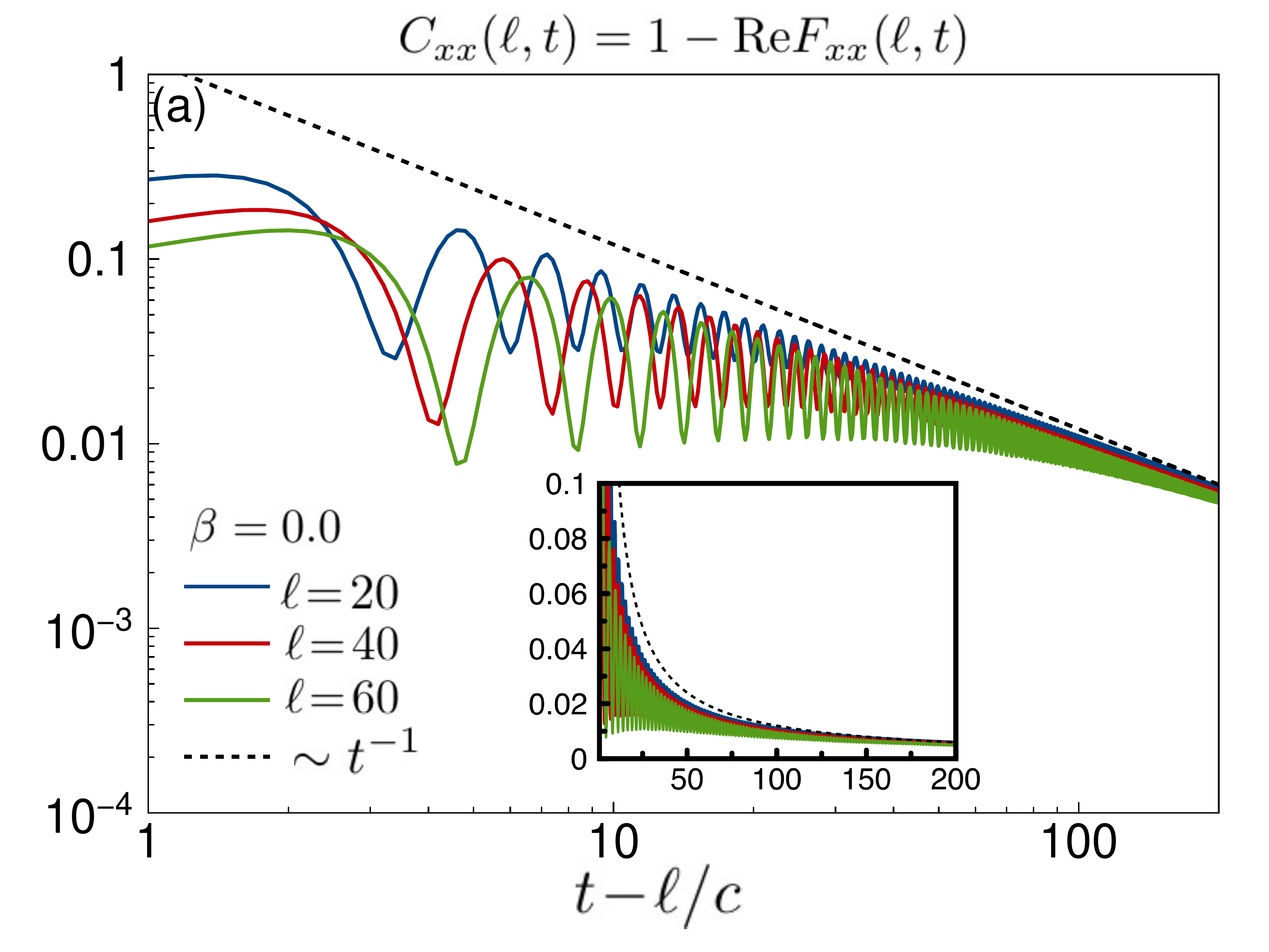}
\includegraphics[width=1\columnwidth]{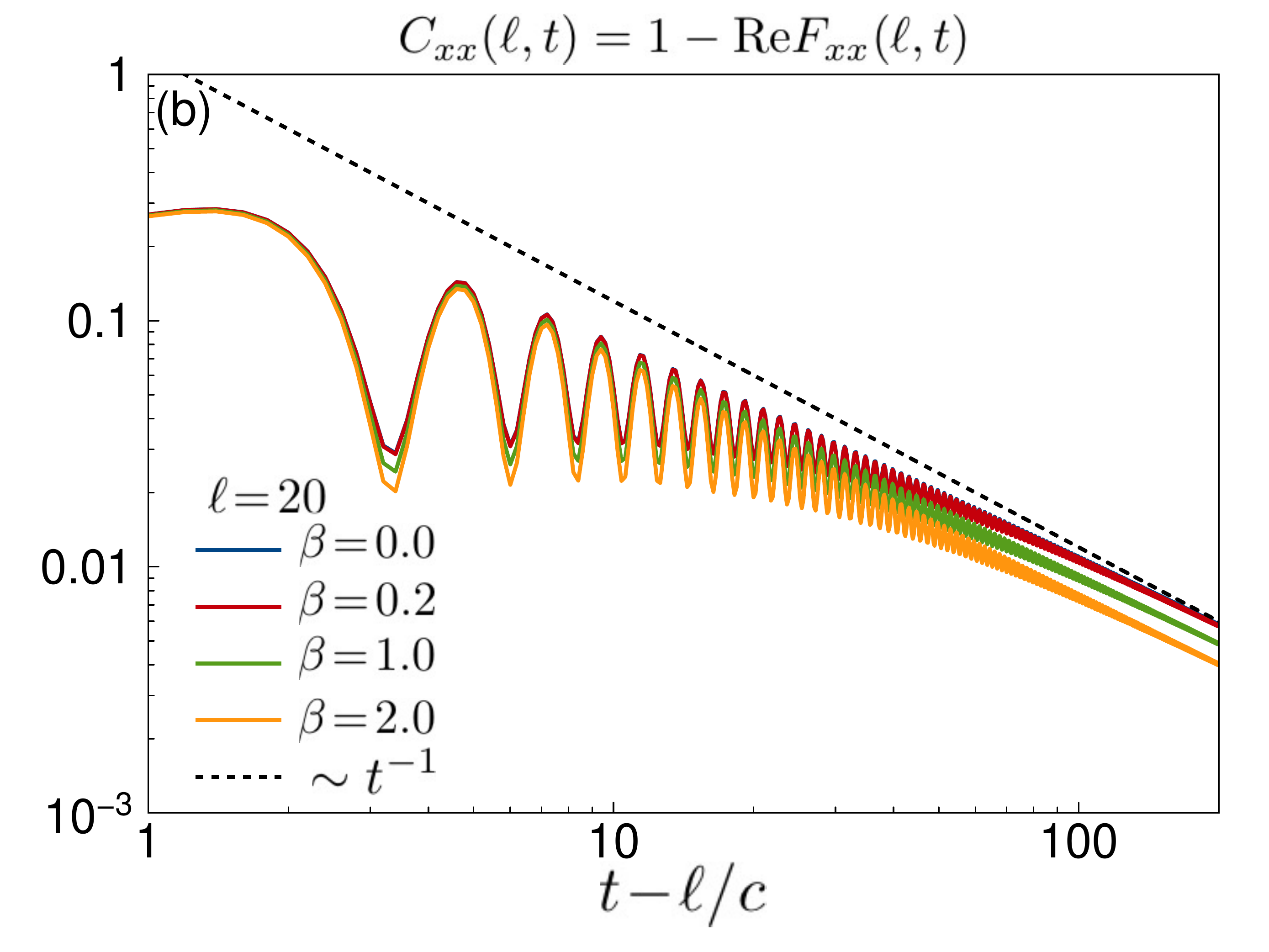}
\caption{\label{fig:xxotoc_long}
Long-time behavior of $C_{xx}(\ell, t)$ in the timelike region; note the log-log scale.
The data is shown as a function of $t$ at fixed $\ell$, where on the horizontal axis we show the time elapsed after the wavefront passes.
Panel (a) shows several different separations $\ell$ and is at infinite temperature; the inset shows the same data on the linear-linear scale.
Panel (b) shows several different temperatures at fixed separation $\ell = 20$.
In all cases, we observe power-law decay $t^{-1}$, which can be understood from the long-time behavior of the fermion correlation functions.
}
\end{figure}

The long time behavior of $C_{xx}(\ell, t)$ is shown in Fig.~\ref{fig:xxotoc_long} for different separations $\ell$ and different inverse temperatures $\beta$.
The data suggests universal $t^{-1}$ behavior independent of $\ell$ and $\beta$.
We can indeed understand this from the stationary phase approximation for the fermionic correlation functions.
The standard stationary phase approximation applied to the fermionic correlation functions gives $t^{-1/2}$ decay at long times.
The full Wick contraction for Eq.~(\ref{eqn:Fxx_in_Majorana}) is complicated but can be obtained by simplifying the calculation of the Pfaffian, see Appendix.~\ref{app:Pf_Fxx} for details.
From this, we can identify the dominant behavior at fixed $\ell$ and long time:
\begin{equation}\label{eqn:Czz_longtime}
C_{zz}(\ell, t) \sim \left(1 - \langle A_0 B_0 \rangle^2 \right) \frac{2}{\pi |\epsilon_\pi''| t} ~,
\end{equation}
where $\epsilon_k''$ is the second derivative of $\epsilon_k$ with respect to $k$ (for $g = 1$ considered here, $|\epsilon_\pi''| = J/2 = 1/2$).
Note that in this expression the temperature dependence enters only in the expectation value $\langle A_0 B_0 \rangle = \langle \sigma_0^x \rangle$, which is zero at infinite temperature and approaches value $0.7698$ at zero temperature (so that the coefficient of the $t^{-1}$ decay is always nonzero).
We can recognize that the $t^{-1}$ decay comes from two pairs of unequal-time contractions and two pairs of equal-time contractions.
Appendix~\ref{app:sigmax_evolution} provides qualitative understanding of this long-time behavior directly from the operator spreading picture.
We also note that the above calculations and qualitative results hold for all $g$ and nonzero temperatures.

It is interesting to compare the OTOC behavior with results for dynamical correlation functions as well as for thermalization of such spin observable in quench settings.
The dynamical correlation function $\langle \sigma_\ell^x(t) \sigma^x_0\rangle = \langle A_\ell(t) B_\ell(t) A_0 B_0 \rangle$ approaches $\langle \sigma_0^x \rangle^2$ in the long-time with $t^{-1}$ power-law.
Indeed, this power law comes from simple calculation, $\langle \sigma_\ell^x(t) \sigma^x_0\rangle - \langle \sigma_0^x \rangle^2 = \langle A_\ell(t) B_0 \rangle \langle B_\ell(t) A_0 \rangle - \langle A_\ell(t) A_0 \rangle \langle B_\ell(t) B_0 \rangle$, and is ultimately related to the long-time behavior of the fermion dynamical correlation function.
However, we note that details of the contraction pieces (i.e., how ``fractions'' of the spin operator get contracted) is different here compared to the OTOC calculation, even though the long-time $t^{-1}$ power law is similar.
Let us now consider quench setting where one starts with some initial state $|\psi_{\text{ini}} \rangle $ (e.g., a product state or a ground state at some other parameter $g' \neq g$) and then evolves under the present Hamiltonian.
Here one finds that $\langle \psi_{\text{ini}} | \sigma_0^x(t) |\psi_{\text{ini}} \rangle$ decays as $t^{-3/2}$ to its equilibrium value in the long-time limit.\cite{essler2016}
Generally, it is clear that the OTOC, dynamical correlation function, and behavior under quench, probe different aspects of the Heisenberg-evolved operator $\sigma_0^x(t)$ 
(see also Appendix~\ref{app:sigmax_evolution}).

\section{ZZ OTOC}\label{sec:zzotoc}
In this section, we discuss the commutator function $C_{zz}(\ell, t) = 1 - \re F_{zz}(\ell, t)$. 
The new feature here is that $\sigma_\ell^z$ is nonlocal in terms of the JW fermions and furthermore changes the fermion parity.
While one can write $\sigma_\ell^z = -(\prod_{j < \ell} A_j B_j) A_\ell$, its Heisenberg evolution $\sigma_\ell^z(t)$ cannot be obtained from the simple free-fermion Heisenberg evolution of the fermions $A_j(t)$ and $B_j(t)$.
The reason is that the original spin Hamiltonian in the fermionic language is in fact composed of projections into two different fermion-parity sectors, with different free-fermion Hamiltonian used in each sector.
The operator $\sigma_\ell^z$, however, changes the fermion-parity, while the Heisenberg evolution of the fermion operators are simple only when working with a fixed free-fermion Hamiltonian over the full Fock space.
Therefore, we need a more sophisticated treatment when calculating the dynamical quantities.

Following McCoy and Abraham,\cite{mccoy1971} we ``double'' the OTOC and consider the following quantity
\begin{equation}
\label{eqn:Gamma_zz}
\Gamma_{zz}(\ell, t; L) \equiv \la \sigma^z_{\frac{L}{2}}(t) \sigma^z_{L\!-\!\ell}(t) \sigma^z_0\sigma^z_{\frac{L}{2}\!-\!\ell} \sigma^z_{\frac{L}{2}}(t) \sigma^z_{L\!-\!\ell}(t) \sigma^z_0\sigma^z_{\frac{L}{2}\!-\!\ell} \ra ~,
\end{equation}
where by periodic boundary conditions site $L - \ell \equiv -\ell$ will be ``close'' to site $0$ (and site $L/2 - \ell$ will be ``close'' to site $L/2$).
Consider large enough $L$ such that $L/2 \gg \ell$ and $L/2 \gg vt$ for some characteristic velocity $v$ (here $v \leq c = 1$).
Invoking the Lieb-Robinson bound and the cluster property,\cite{mccoy1971} we have
\begin{eqnarray}
\Gamma_{zz}(\ell, t; L) &\approx& \la \sigma^z_{\frac{L}{2}}(t) \sigma^z_{\frac{L}{2}\!-\!\ell}\sigma^z_{\frac{L}{2}}(t) \sigma^z_{\frac{L}{2}\!-\!\ell} \ra
\la \sigma^z_{L\!-\!\ell}(t) \sigma^z_0 \sigma^z_{L\!-\!\ell}(t) \sigma^z_0 \ra \nonumber \\
&=& F_{zz}(\ell, t) F_{zz}(-\ell, t) = F_{zz}^2(\ell, t) ~, \label{doubleOTOC}
\end{eqnarray}
where we have used the mirror symmetry $F_{zz}(-\ell, t) = F_{zz}(\ell, t)$.
The advantage of introducing the function $\Gamma_{zz}(\ell, t; L)$ is that $\sigma^z(t)$ operators come in pairs that do not change the fermion parity, which allows expressing the evolution using fixed free-fermion Hamiltonians, so the full function can be calculated via Wick's theorem in terms of the JW fermions.
Again, the evaluations of the Wick's theorem can be conveniently formulated as Pfaffians of appropriate antisymmetric matrices.
We present the details in Appendix~\ref{app:Pf_Fzz}.

Figure~\ref{fig:zzotoc_all} shows $C_{zz}(\ell, t)$ at $g = 1.0$, $\beta = 0$, calculated using the above procedure on a system of size $L = 512$.
Note that since we can only calculate $F_{zz}^2(\ell, t)$, we recover the sign of $\re F_{zz}(\ell, t)$ by requiring ``continuity'' of the ``derivative'' $D_\ell F_{zz}(\ell, t) \equiv F_{zz}(\ell+1, t) - F_{zz}(\ell, t)$ and the known value of $\re F_{zz}(\ell, t) \approx 1$ in the spacelike region $\ell \gg c t$.
We have verified such recovery of the sign also by examining continuity of $\partial_t F_{zz}(\ell, t)$ as we vary $t$.
As in our study of $C_{xx}(\ell, t)$ in Fig.~\ref{fig:XXotoc_all}, we can immediately identify the light cone velocity as the maximal group velocity of the quasiparticles.
On the other hand, we also observe that $C_{zz}(\ell, t)$ approaches a nonzero value inside the light cone at long times.
In fact, in the inset of Fig.~\ref{fig:zzotoc_long}(a), we can see that $\re F_{zz}(\ell, t)$ approaches zero in the long-time limit, and hence $C_{zz}(\ell, t)$ approaches $1$.
Thus $C_{zz}(\ell,t)$ has a ``ball-like'' structure, in contrast to the ``shell-like'' $C_{xx}(\ell, t)$.
We interpret this property of $C_{zz}(\ell, t)$ as a signature of some scrambling of the information in the system.
From the operator spreading point of view, this behavior corresponds to $\sigma_0^z(t)$ having a lot of weight on Pauli-strings with ``random'' $\sigma_\ell^\mu$ in the middle of the strings; more precisely, the infinite-temperature $C_{zz}(\ell, t)$ approaching 1 corresponds to the weight of the strings that have $\sigma_\ell^\mu = \sigma^x$ or $\sigma^y$ approaching 1/2 of the total weight, a kind of ``scrambling.''

\begin{figure}
\includegraphics[width=1\columnwidth]{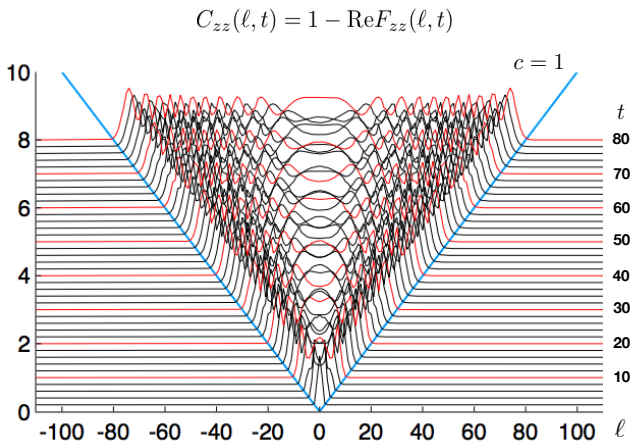}
\caption{\label{fig:zzotoc_all}
(color online) The function $C_{zz}(\ell, t) = 1 - \re F_{zz}(\ell, t)$ for the critical Ising chain ($g = 1$) at infinite temperature ($\beta = 0$), evaluated using the ``doubling trick,'' Eq.~(\ref{doubleOTOC}), on a periodic chain of length $L = 512$.
Here we restore the sign of $\re F_{zz}(\ell, t)$ from $\re \sqrt{\Gamma_{zz}(\ell, t;L)}$ by requiring ``continuity'' of the ``derivative'' $D_\ell \re F_{zz}(\ell, t) = \re F_{zz}(\ell+1, t) - \re F_{zz}(\ell, t)$ (see text for details).
We show data as a function of $\ell$ at fixed $t$, with time steps $\Delta t = 2$.
The traces at fixed $t$ are shifted by $0.1 t$ in the $y$-direction for 3D-like visualization; every $t$ that is multiple of $10$ is marked with red color for easier tracing.
Similarly to $C_{xx}(\ell, t)$ in Fig.~\ref{fig:XXotoc_all}, we can readily identify the light cone and associate it with the maximal quasiparticle velocity $c = 1$.
Unlike $C_{xx}(\ell, t)$, in the timelike region $C_{zz}(\ell, t)$ approaches a nonzero value close to 1 at long times.
In other words, $F_{zz}(\ell, t)$ approaches value close to zero, which suggests scrambling of the information.
}
\end{figure}

\subsection{Early-time behavior of $C_{zz}(\ell, t)$}
\begin{figure}
\includegraphics[width=1\columnwidth]{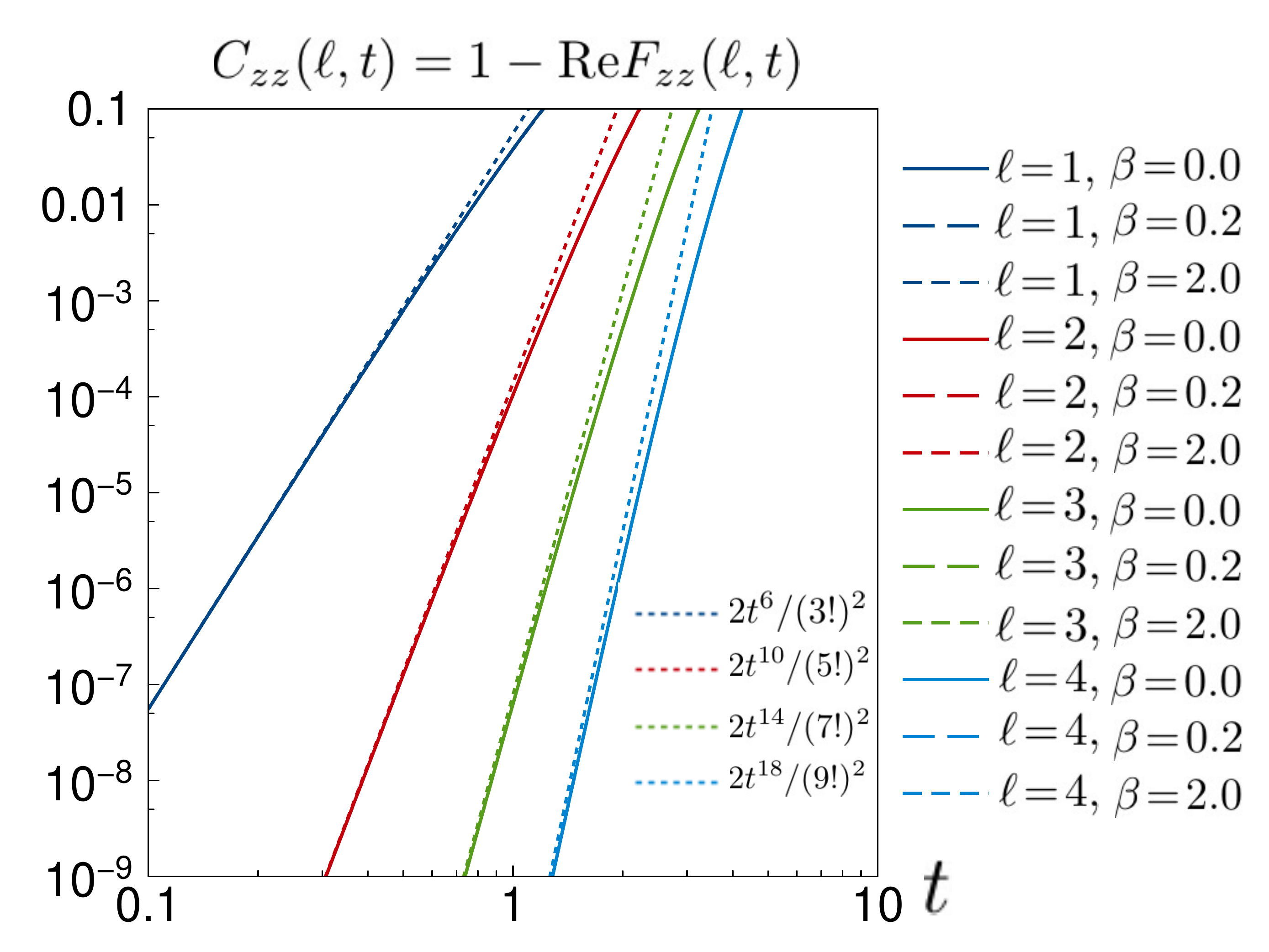}
\caption{\label{fig:zzotoc_shortime}
(color online) The short-time behavior of $C_{zz}(\ell, t)$ examined for several separations $\ell$ and different inverse temperatures $\beta$; the system is the same as in Fig.~\ref{fig:zzotoc_all}.
The early-time growth of $C_{zz}(\ell, t)$ is well described by the ``universal'' power-law given by $\approx 2 t^{4\ell+2}/[(2\ell+1)!]^2$.
}
\end{figure}

The early-time growth of $C_{zz}(\ell, t)$ can be also understood by the argument employing the HBC expansion, Eq.~(\ref{eqn:HBC}).
In this case, for $W = \sigma_0^z$ and $V = \sigma_\ell^z$, the smallest $n$ such that $[L^n(W), V] \neq 0$ is $n = 2\ell + 1$; the corresponding piece in $L^n[W]$ is $-J^{2\ell+1} g^{\ell+1} \sigma_0^x \sigma_1^x \dots \sigma_{\ell-1}^x \sigma_\ell^y$. 
This gives us
\begin{equation}
C_{zz}(\ell, t) \approx 2 \frac{(Jt)^{2(2\ell+1)} g^{2(\ell+1)}}{[(2\ell+1)!]^2} ~.
\end{equation}
In Fig.~\ref{fig:zzotoc_shortime}, we compare the above formula and the numerical results for $C_{zz}(\ell, t)$.
We see that the short-time behavior is well captured by this argument.

\subsection{Behavior of $C_{zz}(\ell, t)$ around the wavefront}
\begin{figure}
\includegraphics[width=1\columnwidth]{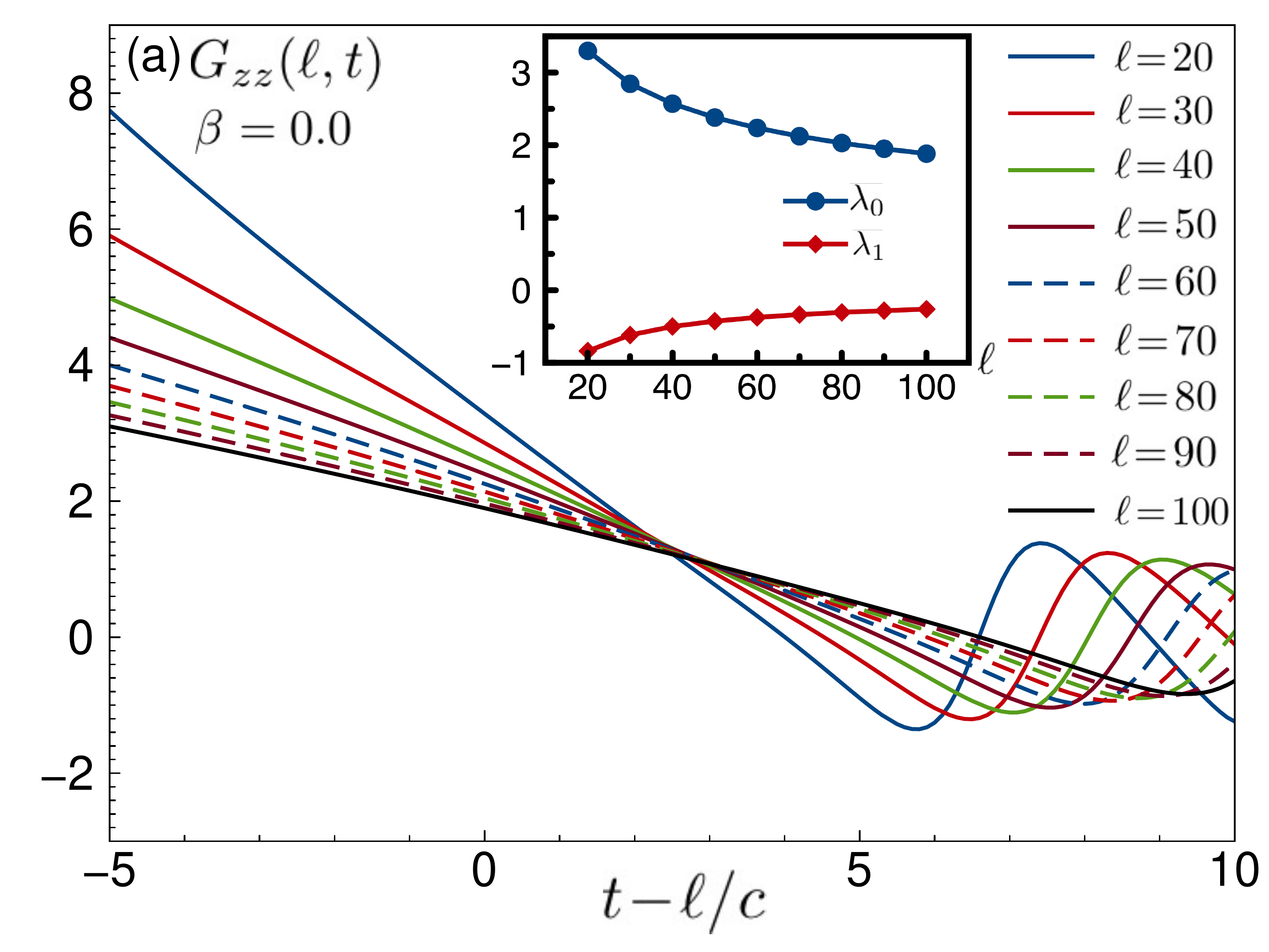}
\includegraphics[width=1\columnwidth]{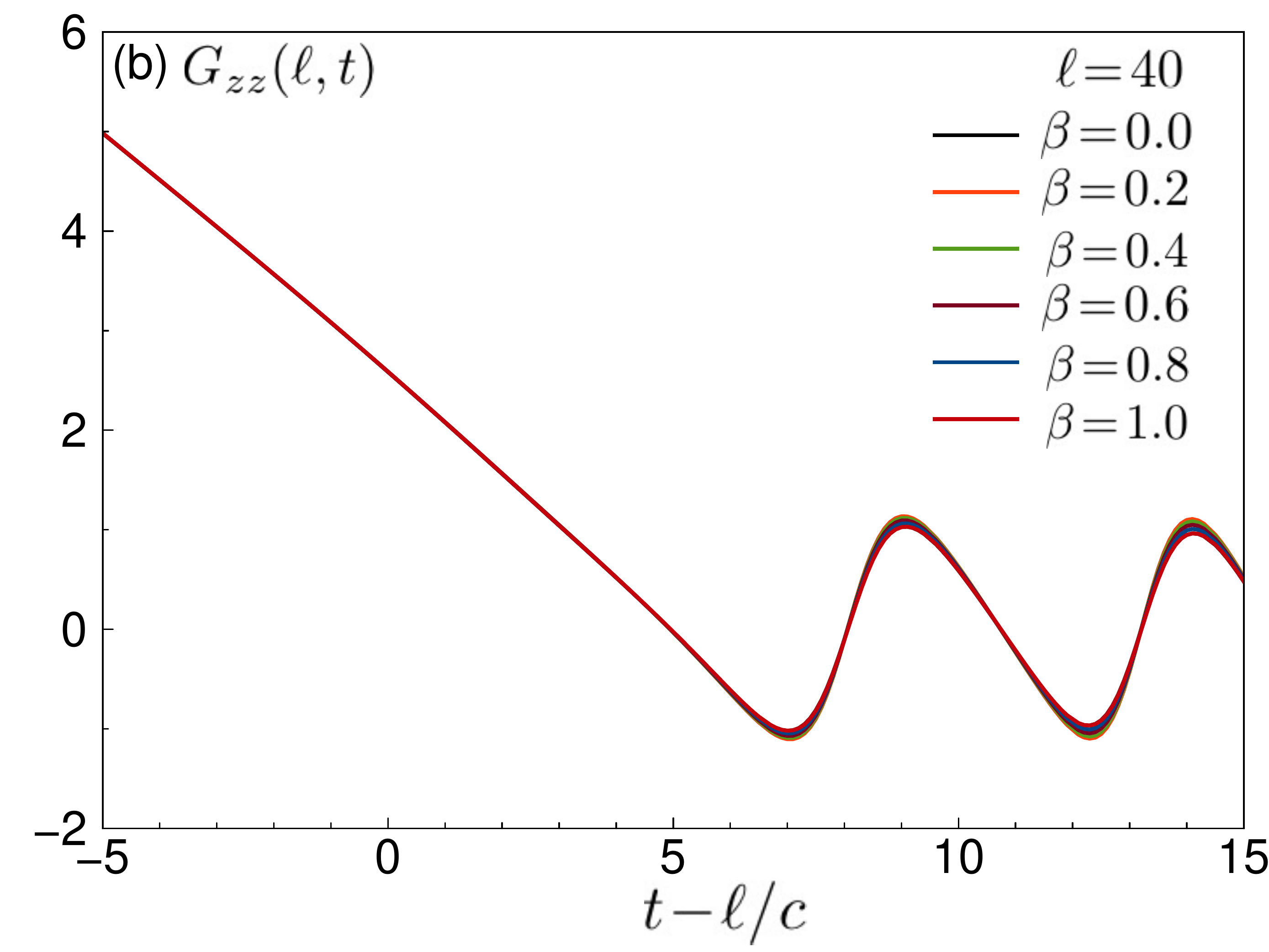}
\caption{\label{fig:zzotoc_wavefront}
The derivative function $G_{zz}(\ell, t) \equiv \partial \ln C_{zz}(\ell, t)/\partial t$ around the wavefront.
(a) $G_{zz}(\ell, t)$ as a function of time for several different separations $\ell$, where the horizontal axis shows time measured relative to the ``exact'' wavefront passage time defined from the known maximal group velocity $c = 1$; this data shows shows strong $\ell$ dependence. 
Around $t - \ell/c = 0$, the behavior of $G_{zz}(\ell, t)$ is well approximated by a linear function.
We fit $G_{zz}(\ell, t)$ to $\lambda_0 + \lambda_1 (t - \ell/c)$ in the region $t - \ell/c \in [-3, 3]$ and show the resulting parameters $g_0$ and $g_1$ for different $\ell$ in the inset.
(b) $G_{zz}(\ell, t)$ for fixed $\ell = 40$ at different inverse temperatures $\beta$; we see that such wavefront characterization does not show strong temperature dependence.
}
\end{figure}

Here we investigate the behavior of $C_{zz}(\ell, t)$ around the wavefront. 
Again, we study the derivative function
\begin{equation}
G_{zz}(\ell, t) \equiv \frac{\partial \ln C_{zz}(\ell, t)}{\partial t} ~.
\end{equation}
Details of the calculation that avoids numerical differentiation are presented in Appendix~\ref{app:Pf_Fzz}.
In principle, if $C_{zz}(\ell, t)$ has the Lyapunov behavior, namely the exponential growth around the wavefront, we should be able to extract this from $G_{zz}(\ell, t=t_0)$, where $t_0 = \ell/c$ is the characteristic wavefront passage time defined using analytically known maximal group velocity $c = 1$.
In Fig.~\ref{fig:zzotoc_wavefront}, we see that $G_{zz}(\ell, t)$ is well described by a linear function $\lambda_0 + \lambda_1 (t - \ell/c)$ around the wave front.
However, the parameters $\lambda_0(\ell)$ and $\lambda_1(\ell)$ have a strong dependence on $\ell$ but very weak dependence on $\beta$.
It is therefore not clear if we should view this functional form as a well defined asymptotic description and identify $\lambda_0$ as the Lyapunov exponent.
One possibility is that when $\ell$ is large, $\lambda_0$ approaches a finite value while $\lambda_1$ approaches zero, therefore it is well-defined when $\ell \rightarrow \infty$ with $\ell/t \sim c$ fixed. 
In this case, around the wavefront, we could say that $C_{zz}(\ell, t) \sim \exp[\lambda_0 (t - \ell/c)]$.
However, we do not seem to have a parametrically large window exhibiting such behavior that could be clearly separated from the short-time and long-time regimes.
Furthermore, any such Lyapunov exponent extracted from our data would be essentially temperature-independent, which would not be consistent with existing proposals.
We do see that the onset of oscillations (which in our mind cuts off any asymptotic description of the wavefront behavior) is pushed to larger $t - \ell/c$ for larger $\ell$, but we do not know if there is any asymptotic functional description to this.
If there is, then similarly to the $C_{xx}$ wavefront in Fig.~\ref{fig:xxotoc_wavefront}, the description should be essentially temperature independent.

\subsection{Unusual slow $t^{-1/4}$ power-law at long time}
\begin{figure}
\includegraphics[width=0.9\columnwidth]{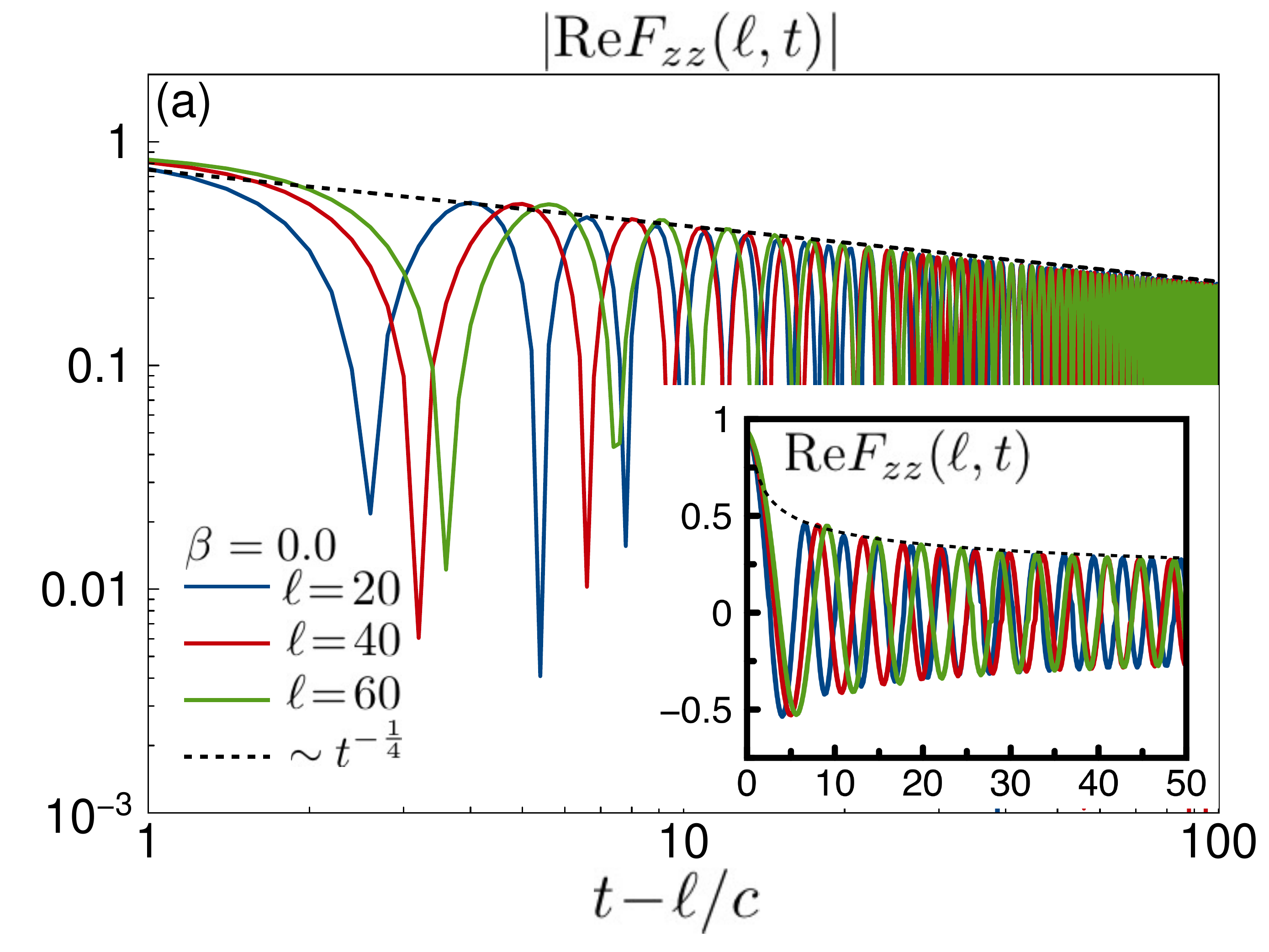}
\includegraphics[width=0.9\columnwidth]{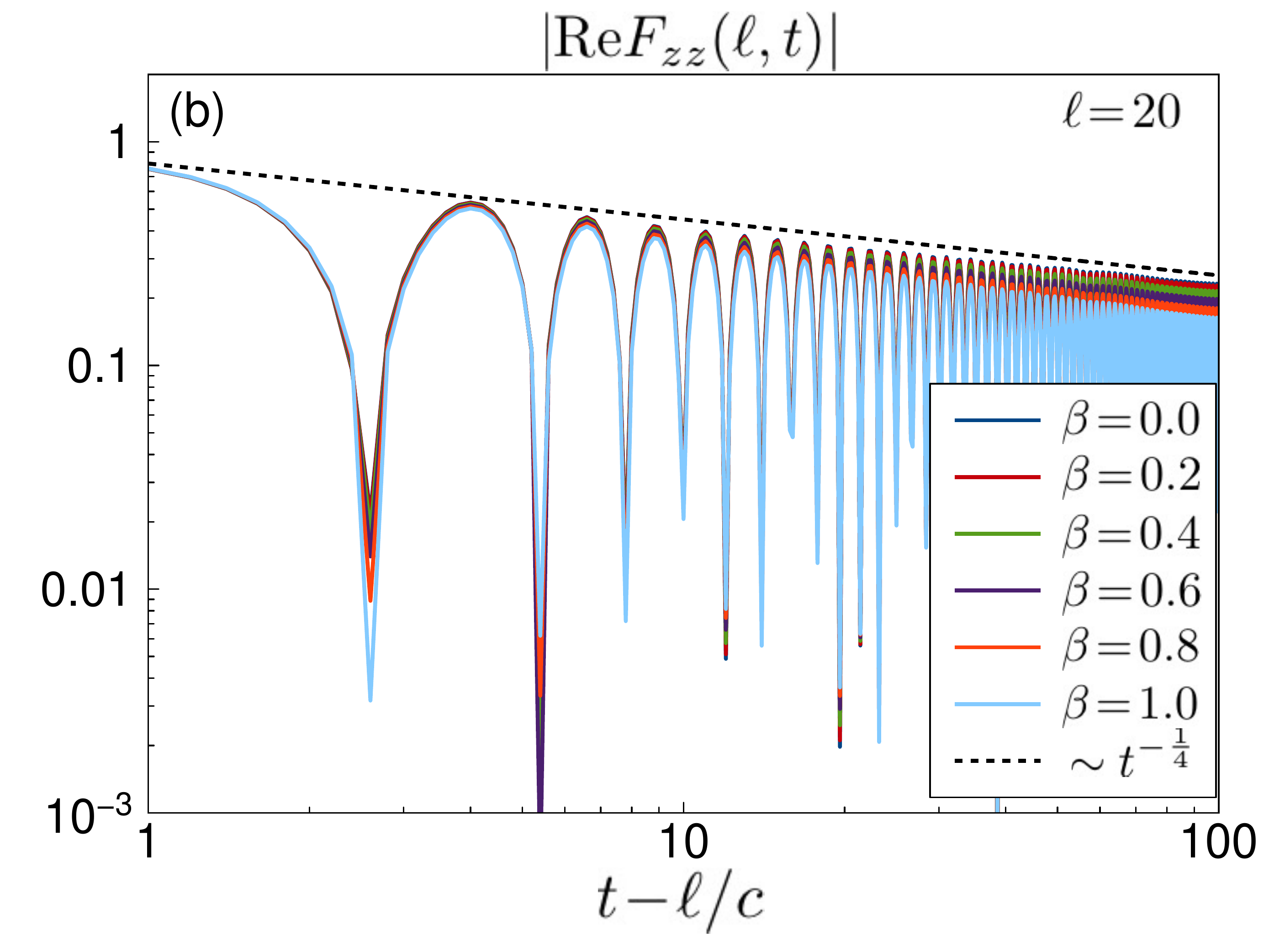}
\includegraphics[width=0.9\columnwidth]{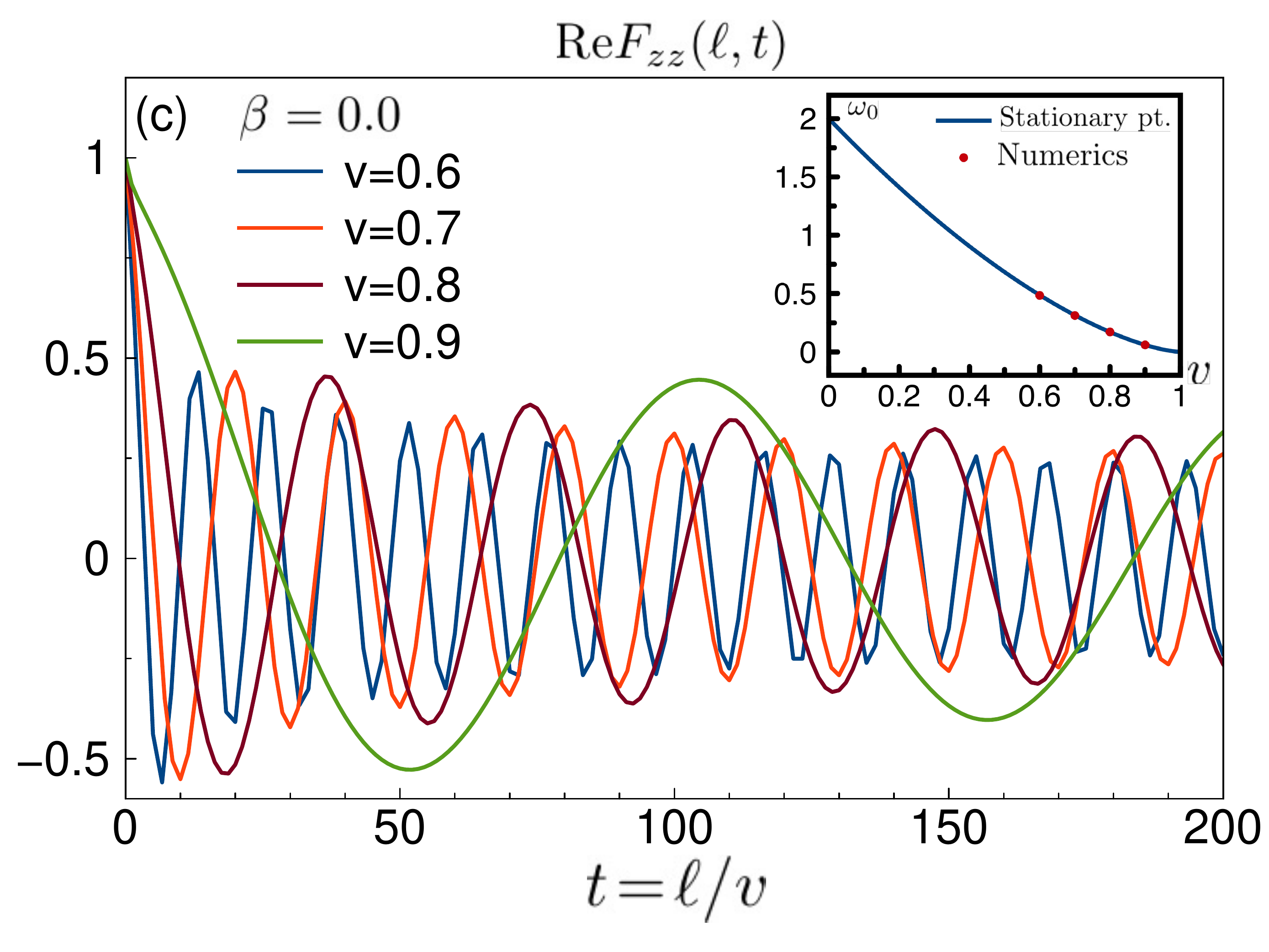}
\caption{\label{fig:zzotoc_long}
(color online) Long-time behavior of $|\re F_{zz}(\ell, t)|$ in the timelike region (i.e., after the wavefront passes) at (a) different separations and (b) different inverse temperatures.
(a) For different separations and $\beta = 0$, in the long-time limit, $|\re F_{zz}(\ell, t)|$ shows $t^{-\frac{1}{4}}$ decay.
Inset: Linear plot of $\re F_{zz}(\ell,t)$, where we fixed the sign by requiring continuity of the derivative $D_t F_{zz}(\ell, t) \equiv F_{zz}(\ell, t + \Delta t) - F_{zz}(\ell, t)$, where $\Delta t$ is the time step in the numerical calculation.
(b) The temperature only affects the coefficient of the power-law decay; in the long-time limit, the decay is still $t^{-1/4}$.
(c) $\re F_{zz}(\ell, t)$ along several different rays $\ell/t = v = \text{const}$ inside the timelike region, where for each $v$ we observe single oscillation frequency that depends on $v$. 
Inset: Comparison of the frequency fitted from from the numerical calculations (red dots) and from the ``stationary phase'' conjecture (blue line) $\omega(v) = \epsilon_{k_0} - k_0 v$, $\partial_k \epsilon|_{k_0} = v$, described in the main text.
}
\end{figure}

An analytical treatment of $C_{zz}(\ell, t)$ is very difficult since it involves analyzing the Pfaffian of a large matrix with essentially infinite dimension in the thermodynamic limit $L \to \infty$.
Here, we analyze it by examining the numerical results in Fig.~\ref{fig:zzotoc_long}.
As before, the data is for the critical Ising chain coupling, $g = 1$, and the calculations are done for system size $L = 512$.
The horizontal axis shows $t\!-\!\ell/c$.
We focus on the long-time behavior of the OTOC $F_{zz}(\ell, t)$ after the wavefront passes. 
We discover that, while $F_{zz}(\ell, t)$ approaches zero in the long-time limit, the approach is described by an oscillating function with a slow power-law envelope $t^{-1/4}$.
This long-time power-law behavior is independent of the separation $\ell$ or the inverse temperature $\beta$.
It is worth mentioning that the finite-temperature calculation for the Ising conformal field theory\cite{roberts2015} gives the same limiting value as our lattice calculation.
However, our $t^{-1/4}$ power-law approach behavior is not described by the conformal field theory.

We can further examine the oscillations by following a specific ray $t = \ell/v$ for varying $v$. 
We show this in Fig.~\ref{fig:zzotoc_long}(c), where we find a single oscillation frequency for each such ray and show its dependence on $v$ in the inset.
We conjecture that the frequency is determined by some ``stationary phase'' approximation on a propagation factor $\exp(i k \ell - i \epsilon_k t)$.
This would give the oscillation frequency as $\omega(v) = \epsilon_{k_0} - k_0 v$, where $k_0$ is the momentum such that the quasiparticle group velocity $\partial \epsilon_k/\partial k |_{k=k_0} = v$.
For $v = 0$, this gives $\omega(v=0) = \epsilon_{k=\pi} = 2$, which is the frequency where the quasiparticle group velocity is zero.
The oscillations in panels Fig.~\ref{fig:zzotoc_long}(a) and \ref{fig:zzotoc_long}(b), where we analyze the limit $t \to \infty$ at fixed $\ell$ which corresponds to $v = 0$, indeed appear to approach this frequency.
However, at present we do not have an analytical understanding of this ``stationary phase'' conjecture and
of the observed $t^{-1/4}$ power law.
We leave this most interesting and mysterious observation as an open question.

In contrast, the dynamical correlation function $\langle \sigma_\ell^z(t) \sigma_0^z \rangle$ decays exponentially in $t$ and $\ell$ as long as the temperature is nonzero.\cite{CAPEL1977211,perk1977,perk1984,its1993,perk2009,sachdev2011} 
The decay length and coherence time depend on the parameter regime ($g$ and $\beta$).
At infinite temperature, the correlation function has a singular behavior $\langle \sigma_\ell^z(t) \sigma_0^z \rangle = \delta_{\ell, 0} e^{-t^2}$,\cite{brandt1976} consistent with vanishing correlation length and coherence time.
Similarly, calculations in quench settings found that $\langle\psi_{\text{ini}} | \sigma_0^z(t) |\psi_{\text{ini}} \rangle$ decays exponentially as well.\cite{calabrese2012,essler2016}
We thus see that there is a qualitative difference between the long-time behaviors of the OTOC and of the dynamical correlations as well as thermalization of the $\sigma^z$ operator.
This indicates that the OTOC captures some different aspects of the physics, and this finding deserved further understanding.

\section{ZX OTOC}\label{sec:zxotoc}
Lastly, we discuss the function $C_{zx}(\ell, t)$.
In the JW fermion language, here we have both a nonlocal operator and a local operator. 
As in the case of $C_{zz}(\ell, t)$, $\sigma^z$ changes the fermion parity sector. 
Therefore, we need to use the ``doubling trick.''
We consider the following function
\begin{eqnarray}
\label{eqn:Gamma_zx}
\Gamma_{zx}(\ell, t; L) &\equiv& \la \sigma^z_{\frac{L}{2}}(t) \sigma^z_{L\!-\!\ell}(t) \sigma^x_0\sigma^x_{\frac{L}{2}\!-\!\ell} \sigma^z_{\frac{L}{2}}(t) \sigma^z_{L\!-\!\ell}(t) \sigma^x_0\sigma^x_{\frac{L}{2}\!-\!\ell} \ra ~. \nonumber
\end{eqnarray}
For large enough system size such that $L/2 \gg \ell$, $L/2 \gg ct$, and using the cluster property and invoking the Lieb-Robinson bound, we have
\begin{eqnarray}
\Gamma_{zx}(\ell, t; L) &\approx& \la \sigma^z_{\frac{L}{2}}(t) \sigma^x_{\frac{L}{2}\!-\!\ell}\sigma^z_{\frac{L}{2}}(t) \sigma^x_{\frac{L}{2}\!-\!\ell} \ra
\la \sigma^z_{L\!-\!\ell}(t) \sigma^x_0 \sigma^z_{L\!-\!\ell}(t) \sigma^x_0 \ra \nonumber \\
&=& F_{zx}(\ell, t) F_{zx}(-\ell, t) = F_{zx}^2(\ell, t) ~.
\end{eqnarray}
In the last line, we have used translational invariance and the mirror symmetry which gives $F_{zx}(-\ell, t) = F_{zx}(\ell, t)$.
We can now express $\Gamma_{zx}(\ell, t; L)$ in terms of the JW fermions evolving under fixed free-fermion Hamiltonians and reduce the calculations to Pfaffians as detailed in Appendix~\ref{app:Pf_Fzx}.

Figure~\ref{fig:xzotoc_all} shows $C_{zx}(\ell, t)$ at $g = 1.0$, $\beta = 0$, calculated using system size $L = 512$.
After the wavefront passes, $C_{zx}(\ell, t)$ approaches a nonzero value in the long-time limit. 
In fact, $\re F_{zx}(\ell, t)$ approaches a negative value.
We identify this behavior as some ``partial scrambling,'' since $\re F_{zx}$ does not approach $1$ (``absence of scrambling'') or $0$ (``total scrambling'').

\begin{figure}
\includegraphics[width=1\columnwidth]{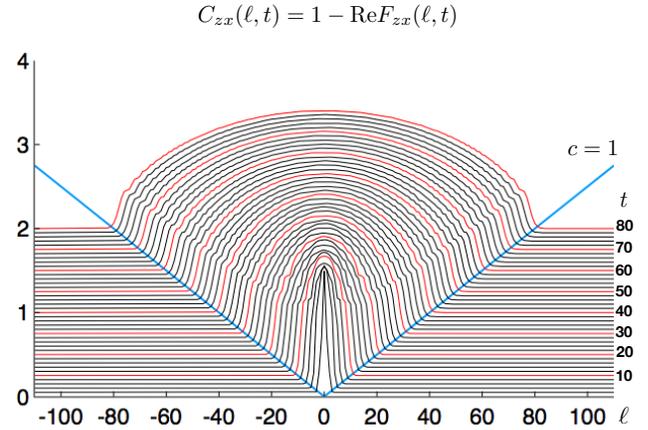}
\caption{\label{fig:xzotoc_all}
(color online) The function $C_{zx}(\ell, t)$ for the same critical Ising chain as in Figs.~\ref{fig:XXotoc_all}~and~\ref{fig:zzotoc_all}.
The traces at fixed $t$ are shifted by $0.025 t$ in the y-direction for 3D-like visualization; every $t$ that is a multiple of $10$ is marked with red color for easier tracing.
We can readily indentify the light cone and the corresponding velocity $c = 1$.
In the timelike region, $C_{zx}(\ell, t)$ approaches a nonzero value larger than 1 in the long-time limit, i.e., $F_{zx}(\ell, t)$ approaches a negative value.
}
\end{figure}

\subsection{Early-time behavior of $C_{zx}(\ell, t)$}
\begin{figure}
\includegraphics[width=1\columnwidth]{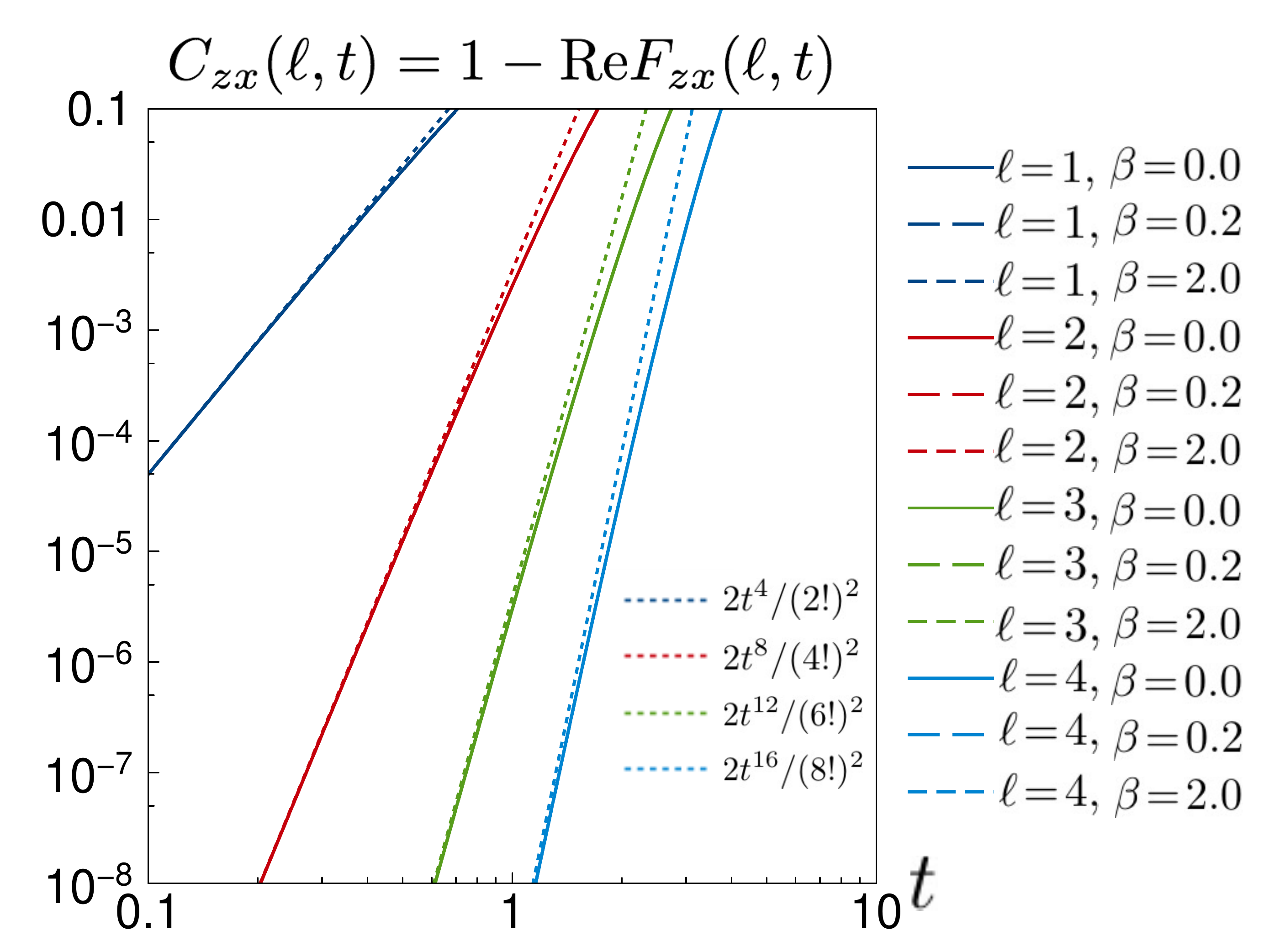}
\caption{\label{fig:xzotoc_short}
(color online) The short-time behavior of $C_{zx}(\ell, t)$ for several separations $\ell$ and different inverse temperatures $\beta$; the system is the same as in Fig.~\ref{fig:xzotoc_all}.
The early-time growth of $C_{zx}(\ell, t)$ is well described by the ``universal'' power law $\approx 2 t^{4\ell}/[(2\ell)!]^2$.
}
\end{figure}

The short-time behavior of $C_{zx}(\ell, t)$ before the wavefront reaches is again described by the ``universal'' power law with position-dependent exponent. 
In this case with $W = \sigma_0^x$ and $V = \sigma_\ell^z$, the smallest $n$ such that $[L^n(W), V] \neq 0$ is $n = 2\ell$, and the corresponding term in $L^n[W]$ is $-J^{2\ell} g^\ell \sigma_0^y \sigma_1^x \dots \sigma_{\ell-1}^x \sigma_\ell^y$.
We thus have the leading behavior 
\begin{equation}
C_{zx}(\ell, t) \approx \frac{2 (Jt)^{4\ell} g^{2\ell}}{[(2\ell)!]^2} ~.
\end{equation}
In Fig.~\ref{fig:xzotoc_short}, we compare the exact numerical results with this leading-order prediction at short time and find good agreement.

\subsection{Behavior of $C_{zx}(\ell, t)$ around the wavefront}
Here we also investigate the behavior of $C_{zx}(\ell, t)$ around the wavefront.
We study the derivative function 
\begin{equation}
G_{zx}(\ell, t) \equiv \frac{\partial \ln C_{zx}(\ell, t)}{\partial t} ~;
\end{equation}
the details of the calculation are presented in Appendix~\ref{app:Pf_Fzx}.
Figure~\ref{fig:xzotoc_wavefront} shows the results around the wavefront defined by $c = 1$.
Similarly to our earlier findings for $G_{xx}(\ell, t)$ and $G_{zz}(\ell, t)$, we see that $G_{zx}(\ell, t)$ has strong $\ell$ dependence but essentially no $\beta$ dependence.
Again, we do not seem to have a parametrically large window around the wavefront that can be sharply separated from the short-time and long-time behaviors, and we definitely do not have any temperature-dependent asymptotic functional description.

\begin{figure}
\includegraphics[width=1\columnwidth]{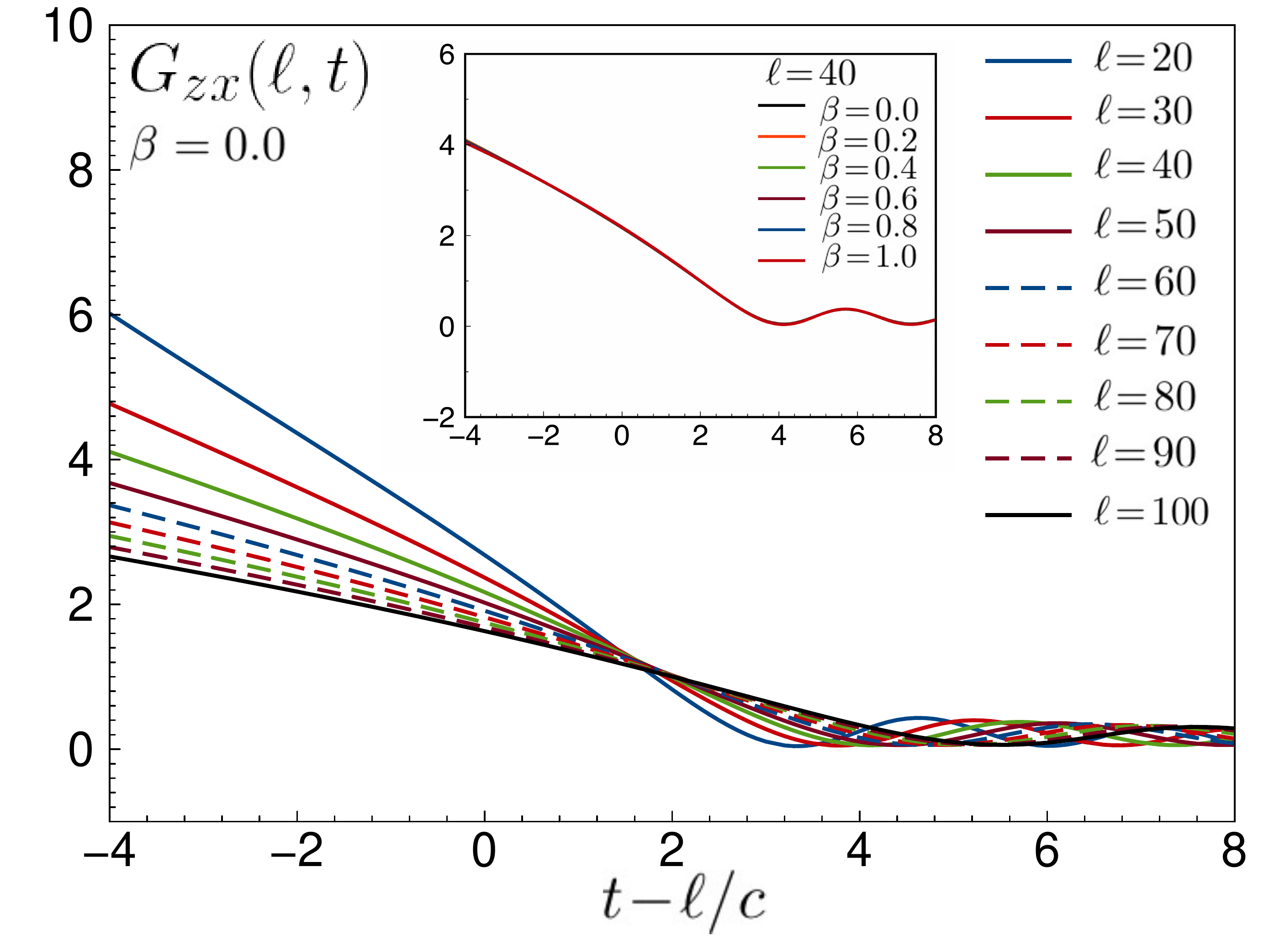}
\caption{\label{fig:xzotoc_wavefront}
The derivative function $G_{zx}(\ell, t) \equiv \partial \ln C_{zx}(\ell, t)/\partial t$ around the wavefront.
$G_{zx}(\ell, t)$ has very strong $\ell$ dependence and no apparent universal description.
Inset: $G_{zx}(\ell, t)$ for fixed $\ell = 40$ and several inverse temperatures $\beta$; there is essentially no temperature dependence.
}
\end{figure}

\subsection{Long-time behavior of $C_{zx}(\ell, t)$}
\begin{figure}
\includegraphics[width=1\columnwidth]{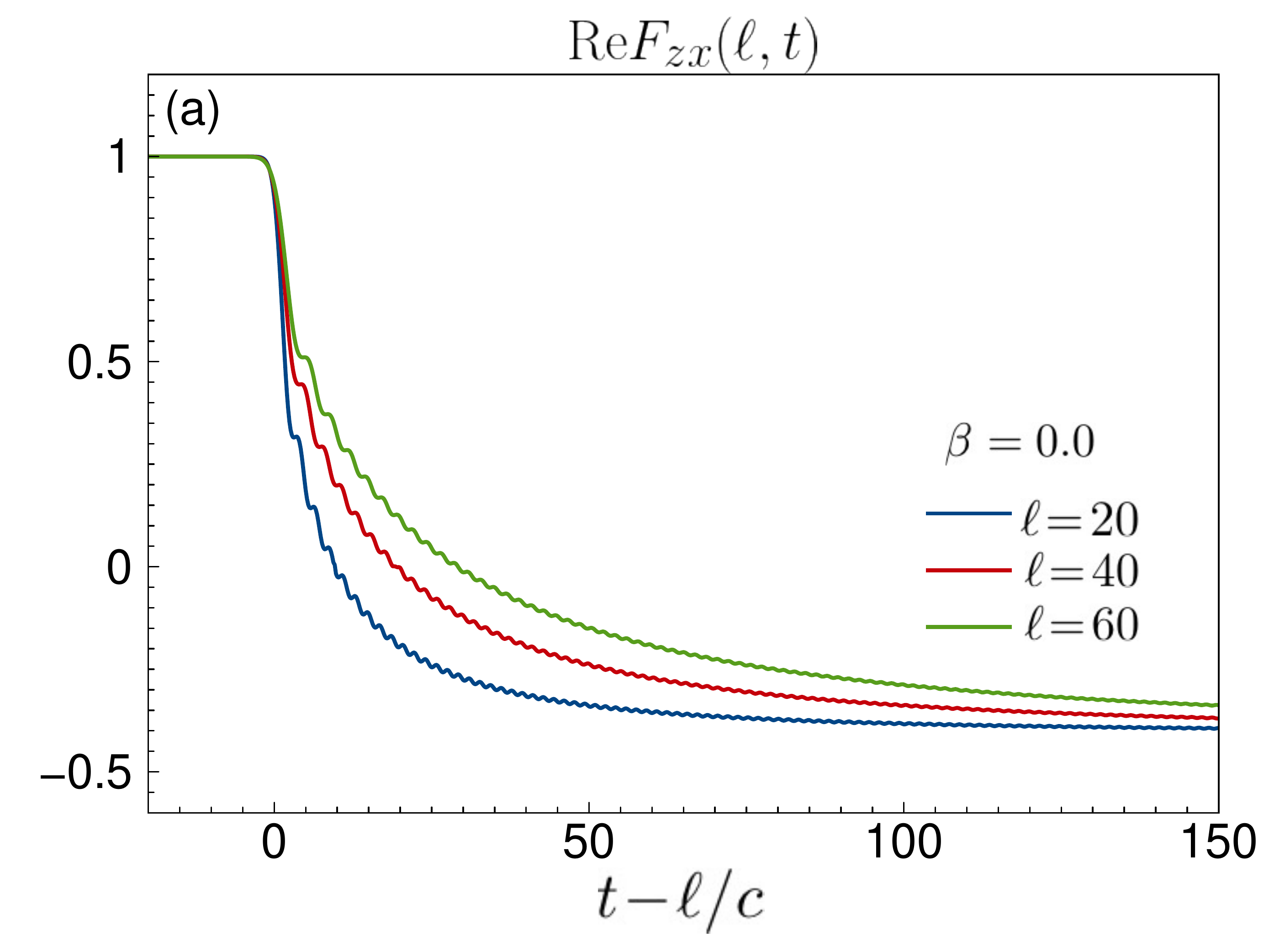}
\includegraphics[width=1\columnwidth]{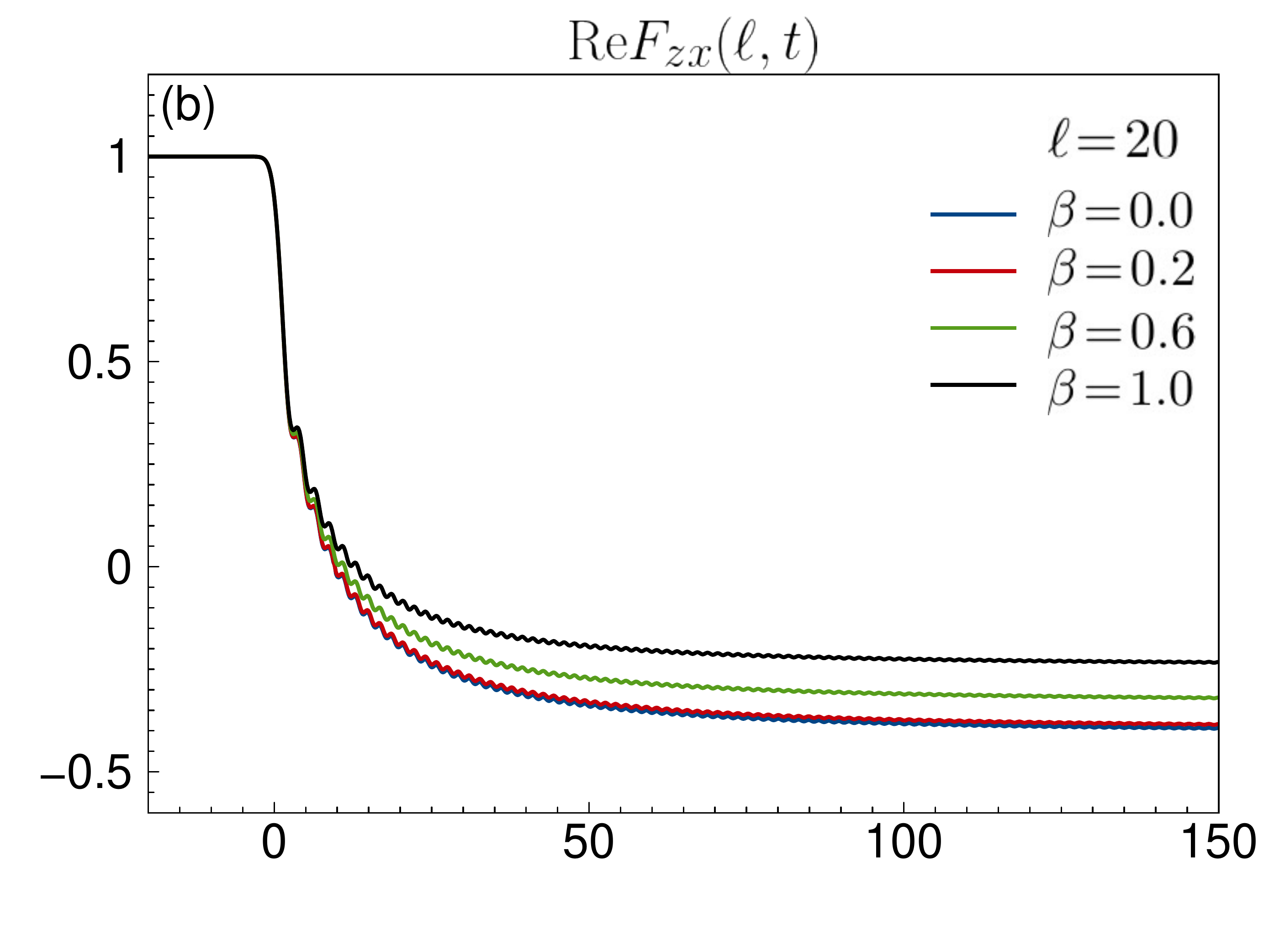}
\caption{\label{fig:xzotoc_long}
(color online)
Long-time behavior of $\re F_{zx}(\ell, t)$ after the wavefront passes.
(a) For different separations $\ell$ at $\beta = 0$, the limiting value as $t \to \infty$ appears to be the same, but the approach behavior has a strong $\ell$ dependence becoming more slow for larger $\ell$.
(b) For different inverse temperatures $\beta$ at fixed $\ell = 20$, we see that both the limiting value and the approach behavior have significant temperature dependence. 
}
\end{figure}

Figure~\ref{fig:xzotoc_long} shows the long-time behavior of the OTOC $F_{zx}(\ell,t)$.
We can see that $F_{zx}$ approaches some nonzero value.
Unlike our results for $F_{xx}$ or $F_{zz}$, the approach of the $F_{zx}$ to the limiting value has a very strong $\ell$ dependence, and we have not been able to identify a ``universal'' long-time description of this behavior.
Furthermore, the limiting value of $F_{zx}(\ell, t)$ when $t \to \infty$ appears to have strong $\beta$ dependence, contributing to our difficulty of finding universal description.

\section{Conclusion}\label{sec:conclusion}
In this paper, we studied the behavior of the OTOC in the integrable quantum Ising model.
We focused on three different OTOCs, which are representative of different combinations of two different types of operators in terms of the JW fermions.
In all cases, we can clearly identify the light cone velocity, which is given by the maximum group velocity of the quasiparticles.
We also argued that before the wavefront reaches, the OTOCs have ``universal'' power-law growth with position-dependent power.
This can be understood from the Hausdorff-Baker-Campbell expansion of the Heisenberg evolution of the operators.
We expect that such early-time power-law growth should also hold in nonintegrable models, as long as one has bounded local Hilbert space and local Hamiltonian.

On the other hand, the long-time behaviors are different for the different OTOC types.
The first type is represented by $C_{xx}(\ell, t)$, which involves only operators that are local in terms of the JW fermions.
The OTOC can be calculated using a finite number of Wick contractions in the fermionic language.
The limiting value of $C_{xx}(\ell, t)$ is zero when $t \to \infty$, which is a hallmark of absence of information scrambling.
The approach is given by $t^{-1}$ power-law at long time, which can be understood from the stationary phase approximation for the fermion correlation function.
This power law persists at any temperature and also at any parameter $g$ of the Ising model.
We expect that OTOCs composed of operators that are local in JW fermions will have similar behavior.

The second type is represented by $C_{zz}(\ell, t)$, which involves only operators that are nonlocal in terms of the JW fermions (these operators contain ``string'' operator when fermionizing the spin model).
Due to this nonlocal character, the OTOC calculation involves $O(L)$ Wick contractions.
In the long-time limit, $F_{zz}(\ell, t)$ approaches zero [$C_{zz}(\ell, t)$ approaches 1], which is a signature of scrambling.
Interestingly, the approach is a very slow power-law $t^{-1/4}$.
While we can tentatively identify the frequency of oscillations that are present in the long-time behavior as coming from the stationary phase approximation, it is not clear how the $t^{-1/4}$ arises. 

The aforementioned $t^{-1/4}$ behavior of $F_{zz}(\ell, t)$ is found at $g = 1$ and any $\beta$.
One immediate question is whether this behavior depends on $g$.
We performed such studies, although the results are not easy to interpret with available system sizes.
For $g > 1$, $F_{zz}(\ell, t)$ appears to approach zero with a faster decay than $t^{-1/4}$.
We observe oscillations with multiple frequencies, which makes it difficult to identify the precise power-law decay.
On the other hand, for $g < 1$, the decay has both oscillating and nonoscillating components, which makes the identification of the long-time behavior even more difficult, but the decay appears to be also faster than $t^{-1/4}$.
Thus, for both $g > 1$ and $g < 1$ we seem to find power-law decay faster than for the ``critical'' coupling $g = 1$.
At present, we do not understand the origin of this qualitative difference, which persists all the way to infinite temperature.
We can only speculate that the full many-body spectrum of the $g = 1$ Ising chain has something special about it compared to $g \neq 1$, even though the thermodynamic phase transition occurs only at zero temperature.

The last type of the OTOC behavior is represented by $C_{zx}(\ell, t)$ and involves both local and nonlocal operators in terms of the JW fermions.
However, the long-time behavior of $C_{zx}(\ell, t)$ has a very strong $\ell$ dependence, while the limiting value also has a $\beta$ dependence.
Because of this, we have not been able to find a ``universal'' ($\ell$-independent) description for $C_{zx}(\ell, t)$ in the long-time limit.

For each of the three types of OTOCs, we tried to study the behavior around the wavefront by considering the time derivative of the logarithm of the corresponding $C_{\mu\nu}(\ell, t)$ function.
In all cases, we found a strong $\ell$ dependence and very weak $\beta$ dependence.
Incidentally, such derivative $G_{zz}(\ell, t)$ can be well described by a linear function around the wavefront, but we do not know if there is some significance to this.
However, we cannot find any parametrically large time window that would enable the exponential-growth description of the wavefront, and we can confidently exclude possibility of any temperature-dependent asymptotic description.

We conclude with some open questions and future directions.
The main unresolved issue in the present paper is finding better physical understanding of the long-time behavior of the $C_{zz}(\ell, t)$ commutator function.
Recent studies\cite{khemani2017} argued that OTOCs in random quantum circuit models with a conserved charge have a power-law approach in the long-time limit due to a diffusive charge spreading.
Our quantum Ising chain, besides the global $Z_2$ symmetry, is also integrable and has many integrals of motion, and it would be interesting to understand if there is a more direct relation between the long-time OTOC behaviors and the integrals of motion.
We would also like to study OTOCs in other spin models that map to free fermions, in particular, with $U(1)$ global symmetry.\cite{lin2017} 
More generally, we would like to understand OTOCs in other integrable models that do not map to free fermions, and also effects of weak integrability breaking.
Another interesting direction is to study integrable models with long-range interactions.
The model in this paper is short-ranged and does not show any Lyapunov growth behavior near the wavefront.
There appears to be mounting evidence that even nonintegrable models but with local interactions and bounded on-site Hilbert spaces do not have a precisely-defined exponential Lyapunov growth regime near the wavefront.
A very recent study\cite{chen2017} proposed that such an exponential growth behavior can be found in nonintegrable models with long-range interactions.
It would be interesting to explore if integrable models with long-range couplings may also exhibit the exponential growth regime.

{\it Note added:} Recently, two papers\cite{xu2018a,khemani2018} appeared that proposed a universal functional form for the OTOC around the wavefront in integrable models, which was conjectured based on the free-fermion calculation for observables that are local in terms of fermions.
This pertains to our discussion of Figs.~\ref{fig:xxotoc_wavefront} and \ref{fig:zzotoc_wavefront}, where we left the possibility of universal description of the wavefront as an open question.
We have actually verified that the proposed wavefront description\cite{khemani2018} indeed holds for both $F_{xx}(\ell, t)$ and $F_{zz}(\ell, t)$, which includes also nonlocal observables that contain the string operator in terms of fermions.
Specifically, on rays with fixed velocity outside the light cone, $(x = vt, t)$ with $v > c$, we have verified that the OTOC has exponential decay $\sim \exp[-\lambda(v) t]$ at long times, with $\lambda(v)$ vanishing as $(v - c)^{3/2}$ as $v \to c$.
Furthermore, the broadening we observed near the wavefront (seen, e.g., in the movement of the first oscillation feature inside the light cone for increasing $\ell$ in  Figs.~\ref{fig:xxotoc_wavefront} and \ref{fig:zzotoc_wavefront}) is also consistent with the proposed broadening $\sim t^{1/3} \sim \ell^{1/3}$.
We thank the authors of Ref.[\onlinecite{khemani2018}] for communications about these points.

\begin{acknowledgments}
The authors would like to thank D.~Aason, D.~Chowdhury, A.~Kitaev, and C.~White for useful discussions.
This work was supported by NSF through Grant No. DMR-1619696 and also by the Institute for Quantum Information and Matter, an NSF Physics Frontiers Center, with support of the Gordon and Betty Moore Foundation.
\end{acknowledgments}

\appendix
\section{Majorana two-point functions and partition sum of the spin model}\label{app:correlation}
Before we proceed, we remind the reader that in this appendix and in Appendices \ref{app:Pf_Fxx}, \ref{app:Pf_Fzz}, and \ref{app:Pf_Fzx}, the time-dependent operators are understood as evolved under the corresponding free-fermion Hamiltonians $H_\NSR$ as explained in the main text after Eq.~(\ref{eqn:secProj}) and determined by the label of the ensemble used, $\langle \dots \rangle_\NSR$.

We now list the Majorana two-point correlation functions, which are ingredients in the applications of the Wick's theorem:
\begin{eqnarray*}
\langle A_m(t) A_n \rangle_\NSR &=& \frac{1}{L} \sum_{k \in K_\NSR} e^{-i k (m-n)} \\
&& \left[\cos(\epsilon_k t) - i \sin(\epsilon_k t) \tanh\left(\frac{\beta \epsilon_k}{2} \right) \right] ~, \\
\langle A_m(t) B_n \rangle_\NSR &=& \frac{1}{L} \sum_{k \in K_\NSR} e^{-i k (m-n)} e^{-i \theta_k} \\
&& \left[\cos(\epsilon_k t) \tanh\left(\frac{\beta \epsilon_k}{2}\right) - i \sin(\epsilon_k t) \right] ~, \\
\langle B_m(t) A_n \rangle_\NSR &=& \frac{-1}{L} \sum_{k \in K_\NSR} e^{-i k (m-n)} e^{i \theta_k}\\
&& \left[\cos(\epsilon_k t) \tanh\left(\frac{\beta \epsilon_k}{2}\right) - i \sin(\epsilon_k t) \right] ~, \\
\langle B_m(t) B_n \rangle_\NSR &=& \frac{-1}{L} \sum_{k \in K_\NSR} e^{-i k (m-n)} \\
&& \left[\cos(\epsilon_k t) - i \sin(\epsilon_k t) \tanh\left(\frac{\beta \epsilon_k}{2}\right) \right] ~. 
\end{eqnarray*}
The equal-time correlations are thus 
$\langle A_m A_n \rangle_\NSR = -\langle B_m B_n \rangle_\NSR = \delta_{mn}$,
$\langle A_m B_n \rangle_\NSR = \frac{1}{L} \sum_{k \in K_\NSR} e^{-i k (m-n)} e^{-i \theta_k} \tanh(\beta \epsilon_k/2)$, and 
$\langle B_m A_n \rangle_\NSR = -\langle A_n B_m \rangle_\NSR$.

We define matrices $[{\tt AA_\NSR}](t)$ with matrix elements 
$[{\tt AA_\NSR}]^m_n(t) \equiv \langle A_m(t) A_n \rangle_\NSR$ and analogously for $[{\tt AB_\NSR}]^m_n(t)$, $[{\tt BA_\NSR}]^m_n(t)$ and $[{\tt BB_\NSR}]^m_n(t)$.
We also use $[{\tt I}]$ and $[{\tt 0}]$ to denote identity and zero matrices.
For simplicity, the equal-time correlators are denoted by omitting the time argument.
We also use $[{\tt AA_\NSR}]^{m=i:j}_{n=k:l}(t)$ to represent the submatrix of $[{\tt AA_\NSR}](t)$ with row index from $i$ to $j$ and column index from $k$ to $l$.
We will frequently omit $\NSR$ in $[{\tt AA_\NSR}](t)$ and other matrices since it will be clear from the context which matrix is used.

As we mentioned in the main text, the calculation of the partition function $Z$ of the spin model is less straightforward as it involves the projectors. 
Specifically, we have
\begin{equation}
Z = \text{Tr}(e^{-\beta H_\text{NS}} P_+) + \text{Tr}(e^{-\beta H_\text{R}} P_-) ~,
\end{equation}
where $P_\pm = [1 \pm (-1)^{N_\text{tot}}]/2$.
We therefore have 
\begin{equation}
Z = Z_\text{NS} \frac{1 + \langle (-1)^{N_\text{tot}} \rangle_\text{NS}}{2} 
+ Z_\text{R} \frac{1 - \langle (-1)^{N_\text{tot}} \rangle_\text{R}}{2} ~.
\end{equation}
In the Majorana fermion language, 
\begin{equation}
(-1)^{N_\text{tot}} = (-1)^{L(L-1)/2} \prod_{j=0}^{L-1} A_j \prod_{j=0}^{L-1} B_j ~.
\end{equation}
Defining $D_\NSR = [{\tt AB_\NSR}]^{m=0:L-1}_{n=0:L-1}$ and
\begin{equation}
F_\NSR = \begin{pmatrix} 
0 & D_\NSR \\
-D_\NSR^\text{T} & 0 
\end{pmatrix} ~,
\end{equation}
we have by Wick's theorem (recalling that equal-time contractions $\langle A_m A_n \rangle_\NSR$ and $\langle B_m B_n \rangle_\NSR$ vanish for $m \neq n$):
\begin{eqnarray}
\langle (-1)^{N_\text{tot}} \rangle_\NSR &=& (-1)^{L(L-1)/2} \text{Pf}(F_\NSR)\nonumber \\
&=& \det(D_\NSR) ~.
\end{eqnarray}

\section{Pfaffian calculation of $F_{xx}(\ell, t)$}\label{app:Pf_Fxx}
In this appendix, we present details of the calculation of $F_{xx}(\ell, t)$ for the spin chain with periodic boundary conditions using the Pfaffian method.
We define $2 \times 2$ matrices 
\begin{eqnarray*}
R^{xx}_\NSR &=& \begin{pmatrix}
0 & [{\tt AB}]^0_0 \\
-[{\tt AB}]^0_0 & 0  
\end{pmatrix} ~, \\
S^{xx}_\NSR &=& \begin{pmatrix}
[{\tt AA}]^\ell_0(t) & [{\tt AB}]^\ell_0(t) \\
[{\tt BA}]^\ell_0(t) & [{\tt BB}]^\ell_0(t) 
\end{pmatrix} ~, \\
U^{xx}_\NSR &=& \begin{pmatrix}
[{\tt AA}]^0_\ell(-t) & [{\tt BA}]^0_\ell(-t) \\
[{\tt AB}]^0_\ell(-t) & [{\tt BB}]^0_\ell(-t)
\end{pmatrix} ~, \\
J^{xx} &=& \begin{pmatrix}
1 & 0 \\
0 & -1 
\end{pmatrix} ~; 
\end{eqnarray*}
$4 \times 4$ matrices
\begin{eqnarray*}
M^{xx}_\NSR &=& \begin{pmatrix} 
R^{xx}_\NSR & S^{xx}_\NSR \\
-(S^{xx}_\NSR)^\text{T} & R^{xx}_\NSR
\end{pmatrix} ~, \nonumber \\
N^{xx}_\NSR &=& \begin{pmatrix} 
J^{xx} \!+\! R^{xx}_\NSR & S^{xx}_\NSR \\
(U^{xx}_\NSR)^\text{T} & J^{xx} \!+\! R^{xx}_\NSR
\end{pmatrix} ~;
\end{eqnarray*}
$8 \times 8$ matrix
\begin{eqnarray*}
\Phi^{xx}_\NSR &=& \begin{pmatrix}
M^{xx}_\NSR & N^{xx}_\NSR \\
-(N_\NSR^{xx})^\text{T} & M^{xx}_\NSR
\end{pmatrix} ~;
\end{eqnarray*}
and $4 \times 2L$ matrix
\begin{eqnarray*}
Q^{xx}_{\NSR} &=&
\begin{pmatrix}
[{\tt AA}]^\ell_{n=0:L-1}(t) & [{\tt AB}]^\ell_{n=0:L-1}(t) \\
[{\tt BA}]^\ell_{n=0:L-1}(t) & [{\tt BB}]^\ell_{n=0:L-1}(t) \\
[{\tt I}]^0_{n=0:L-1} & [{\tt 0}]^0_{n=0:L-1} \\
[{\tt 0}]^0_{n=0:L-1} & -[{\tt I}]^0_{n=0:L-1}
\end{pmatrix} ~.
\end{eqnarray*}
Applying Wick's theorem, we have 
\begin{eqnarray*}
&& \langle \sigma_\ell^x(t) \sigma_0^x \sigma_\ell^x(t) \sigma_0^x \rangle_\NSR
= \text{Pf}(\Phi^{xx}_\NSR) ~, \\
&& \langle \sigma_\ell^x(t) \sigma_0^x \sigma_\ell^x(t) \sigma_0^x (-1)^{N_\text{tot}} \rangle_\NSR \nonumber\\
&&=(-1)^{\frac{L(L-1)}{2}}\text{Pf}
\begin{pmatrix}
M^{xx}_{\NSR} & N^{xx}_{\NSR} &Q^{xx}_{\NSR} \\
-(N_{\NSR}^{xx})^{\text{T}} & M^{xx}_{\NSR} &Q^{xx}_{\NSR} \\
-(Q^{xx}_{\NSR})^{\text{T}} & -(Q^{xx}_{\NSR})^{\text{T}} & F_{\NSR}
\end{pmatrix} ~. \nonumber \\
\end{eqnarray*}
In the thermodynamic limit, we expect 
$\langle \sigma_\ell^x(t) \sigma_0^x \sigma_\ell^x(t) \sigma_0^x \rangle = \langle \sigma_\ell^x(t) \sigma_0^x \sigma_\ell^x(t) \sigma_0^x \rangle_\NSR$.

To obtain a compact analytical result, we focus on $\langle \sigma_\ell^x(t) \sigma_0^x \sigma_\ell^x(t) \sigma_0^x \rangle_\text{NS}$.
The Pfaffian can be simplified as follows (we omit the labels ``$xx$'' and ``$\text{NS}$'' for brevity):
\begin{eqnarray*}
&& \text{Pf}\begin{pmatrix}
R & S & J\!+\!R & S \\
-S^\text{T} & R & U^\text{T} & J\!+\!R \\
-J\!+\!R & -U & R & S \\
-S^\text{T} & -J\!+\!R & -S^\text{T} & R \\ 
\end{pmatrix} \\
&& = \text{Pf}\begin{pmatrix}
R & S & J\!+\!R & 0 \\
-S^\text{T} & R & U^\text{T} & J \\
-J\!+\!R & -U & R & S\!+\!U \\
0 & -J & -(S^\text{T}\!+\!U^\text{T}) & 0\\ 
\end{pmatrix} \\
&& = \text{Pf}\begin{pmatrix}
R & S & J & 0 \\
-S^\text{T} & R & S^\text{T}\!+\!U^\text{T} & J \\
-J & -(S\!+\!U) & 0 & S\!+\!U\\
0 & -J & -(S^\text{T}\!+\!U^\text{T}) & 0\\ 
\end{pmatrix} \\
&& = \text{Pf}\begin{pmatrix}
R & S & J & 0 \\
-S^{\text{T}} & R & 0 & J \\
-J & 0 & 0 & S\!+\!U\\
0 & -J & -(S^{\text{T}}\!+\!U^{\text{T}}) & 0\\ 
\end{pmatrix} ~.
\end{eqnarray*}
The matrix $S + U$ is 
\begin{equation*}
S + U = 2 \begin{pmatrix}
\text{Re}\langle A_\ell(t) A_0 \rangle & i\, \text{Im}\langle A_\ell(t) B_0 \rangle \\
i\, \text{Im}\langle B_\ell(t) A_0 \rangle & \text{Re}\langle B_\ell(t) B_0 \rangle 
\end{pmatrix} ~,
\end{equation*} 
and we therefore obtain
\begin{eqnarray*}
F_{xx}(\ell, t) &=& 1 + 2 i\, \la B_\ell(t) A_0 \ra \, \im \la B_\ell(t) A_0 \ra \\
&+& 2 i\, \la A_\ell(t) B_0 \ra \, \im \la A_\ell(t) B_0 \ra \\
&-& 2 \la A_\ell(t) A_0 \ra \, \re \la A_\ell(t) A_0 \ra \\
&-& 2 \la B_\ell(t) B_0 \ra \, \re \la B_\ell(t) B_0 \ra \\
&-& 4 \la A_0 B_0 \ra^2 \, \re \la A_\ell(t) A_0 \ra \, \re \la B_\ell(t) B_0 \ra \\
&+& 4 \la A_\ell(t) A_0 \ra \, \la B_\ell(t) B_0 \ra \, \im \la A_\ell(t) B_0 \ra \, \im \la B_\ell(t) A_0 \ra \\
&+& 4 \la A_\ell(t) A_0 \ra \, \la B_\ell(t) B_0 \ra \, \re \la A_\ell(t) A_0 \ra \, \re \la B_\ell(t) B_0 \ra \\
&-& 4 \la A_\ell(t) B_0 \ra \, \la B_\ell(t) A_0 \ra \, \im \la A_\ell(t) B_0 \ra \, \im \la B_\ell(t) A_0 \ra \\
&-& 4 \la A_\ell(t) B_0 \ra \, \la B_\ell(t) A_0 \ra \, \re \la A_\ell(t) A_0 \ra \, \re \la B_\ell(t) B_0 \ra ~.
\end{eqnarray*}

We can identify the leading behavior as 
\begin{eqnarray}
\re F_{xx}(\ell, t) &\sim& 1 - 2 \left[(\im \la B_\ell(t)A_0 \ra)^2+(\im \la A_\ell(t)B_0 \ra)^2 \right. \nonumber \\
&& ~~~~~~~ \left. + (\re \la A_\ell(t) A_0 \ra)^2 + (\re \la B_\ell(t) B_0 \ra)^2 \right] \nonumber \\
&-& 4 \la A_0 B_0 \ra^2 \left[ \im \la A_\ell(t) B_0 \ra \, \im \la B_\ell(t) A_0 \ra \right. \nonumber \\
&& ~~~~~~~~~~~~~ \left. + \re \la A_\ell(t) A_0 \ra \, \re \la B_\ell(t) B_0 \ra \right] ~.\nonumber \\
\end{eqnarray}

If we follow the ray $\ell/t = v$, where $|v| < c$, in the long-time limit, we can use the stationary phase approximation and obtain
\begin{eqnarray*}
\re \la A_\ell(t) A_0 \ra &\sim& \sqrt{\frac{1}{2\pi \epsilon''_{k_0} t}} \cos\left( \omega_0 t - \frac{\pi}{4} \right) ~, \\
\re \la B_\ell(t) B_0 \ra &\sim& -\sqrt{\frac{1}{2\pi \epsilon''_{k_0} t}} \cos\left( \omega_0 t - \frac{\pi}{4} \right) ~, \\
\im \la A_\ell(t) B_0 \ra &\sim& -\sqrt{\frac{1}{2\pi \epsilon''_{k_0} t}} \sin\left( \omega_0 t - \theta_{k_0} - \frac{\pi}{4} \right) ~, \\
\im \la B_\ell(t) A_0 \ra &\sim& \sqrt{\frac{1}{2\pi \epsilon''_{k_0} t}} \sin\left( \omega_0 t + \theta_{k_0} - \frac{\pi}{4} \right) ~.
\end{eqnarray*}
Here $k_0$ is the wavevector satisfying $d\epsilon_k/dk|_{k_0} = v$, and $\omega_0 \equiv \epsilon_{k_0} - k_0 v$ is the frequency.
In particular, if we fix $\ell$ and consider long-time limit, this effectively corresponds to $v = 0$ and gives $k_0 = \pi$ and $\theta_\pi = 0$.
In this case we find that the limiting behavior of $\re F_{xx}(\ell, t)$ is $t^{-1}$ decay without oscillation, and we obtain Eq.~(\ref{eqn:Czz_longtime}) quoted in the main text. 

To calculate $G_{xx}(\ell, t) \equiv \frac{\partial \ln C_{xx}(\ell, t)}{\partial t}$, we use 
\begin{equation}
G_{xx}(\ell, t) = \frac{-1}{1 - \re{\text{Pf}[\Phi^{xx}_\text{NS}]}} \re \left(\frac{d\text{Pf}[\Phi^{xx}_\text{NS}]}{dt} \right) ~,
\end{equation}
where the derivative of the Pfaffian can be calculated as 
\begin{equation}
\frac{d\text{Pf}[\Phi^{xx}_\text{NS}]}{dt} = \frac{1}{2}\text{Pf}[\Phi^{xx}_\text{NS}] \, \text{Tr}\left[(\Phi^{xx}_\text{NS})^{-1} \frac{d\Phi^{xx}_\text{NS}}{dt} \right] ~.
\end{equation}
The difference between results obtained using ``$\text{NS}$'' and ``$\text{R}$'' boundary conditions is very small for large enough systems.

\section{Pfaffian calculation of $F_{zz}(\ell, t)$}\label{app:Pf_Fzz}
Here we present details of the calculation of $F_{zz}(\ell, t)$.
The ``doubled'' OTOC $\Gamma_{zz}(\ell, t; L)$, Eq.~(\ref{eqn:Gamma_zz}), can be written in terms of the JW fermions as 
\begin{eqnarray*}
\Gamma_{zz}(\ell, t; L) &=& \left\la \left(\prod_{j=\frac{L}{2}}^{L\!-\!\ell\!-\!1} B_{j}(t) A_{\!j\!+\!1}(t) \right) \left(\prod_{j=0}^{\frac{L}{2}\!-\!\ell\!-\!1} B_j A_{j\!+\!1} \right) \right. \\
&& \left. \left(\prod_{j=\frac{L}{2}}^{L\!-\!\ell\!-\!1} B_{j}(t) A_{j\!+\!1}(t) \right) \left(\prod_{j=0}^{\frac{L}{2}\!-\!\ell\!-\!1} B_j A_{j\!+\!1} \right)\right\ra ~.
\end{eqnarray*}
We again need to calculate both ``$\text{RS}$'' and ``$\text{N}$'' pieces.
We define $(L - 2\ell) \times (L - 2\ell)$ matrices
\begin{eqnarray*}
R^{zz}_\NSR &=& \begin{pmatrix}
[0] & [{\tt AB}]^{m=1:\frac{L}{2}\!-\!\ell}_{n=0:\frac{L}{2}\!-\!\ell\!-\!1} \\
[{\tt BA}]^{m=0:\frac{L}{2}\!-\!\ell\!-\!1}_{n=1:\frac{L}{2}\!-\!\ell} & [0]
\end{pmatrix} ~, \\
S^{zz}_\NSR &=& \begin{pmatrix}
[{\tt AA}]^{m=\frac{L}{2}:L\!-\!\ell\!-\!1}_{n=0:\frac{L}{2}\!-\!\ell\!-\!1}(t) & [{\tt AB}]^{m=1+\frac{L}{2}:L\!-\!\ell}_{n=0:\frac{L}{2}\!-\!\ell\!-\!1}(t) \\
[{\tt BA}]^{m=\frac{L}{2}:L\!-\!\ell\!-\!1}_{n=1:\frac{L}{2}\!-\!\ell}(t) & [{\tt BB}]^{m=\frac{L}{2}:L\!-\!\ell\!-\!1}_{n=0:\frac{L}{2}\!-\!\ell\!-\!1}(t) 
\end{pmatrix} ~, \\
U^{zz}_\NSR &=& \begin{pmatrix}
[{\tt AA}]^{m=\frac{L}{2}:L\!-\!\ell\!-\!1}_{n=0:\frac{L}{2}\!-\!\ell\!-\!1}(-t) & [{\tt BA}]^{m=1+\frac{L}{2}:L\!-\!\ell}_{n=0:\frac{L}{2}\!-\!\ell\!-\!1}(-t) \\
[{\tt AB}]^{m=\frac{L}{2}:L\!-\!\ell\!-\!1}_{n=1:\frac{L}{2}\!-\!\ell}(-t) &[{\tt BB}]^{m=\frac{L}{2}:L\!-\!\ell\!-\!1}_{n=0:\frac{L}{2}\!-\!\ell\!-\!1}(-t)
\end{pmatrix} ~, \\
J^{zz} &=& \begin{pmatrix}
[{\tt I}] & [{\tt 0}] \\
[{\tt 0}] & -[{\tt I}]
\end{pmatrix} ~,
\end{eqnarray*}
where $[{\tt I}]$ and $[{\tt 0}]$ are $(\frac{L}{2}\!-\!\ell) \times (\frac{L}{2}\!-\!\ell)$ unit and zero matrices respectively.
We then construct $2(L - 2\ell) \times 2(L - 2\ell)$ matrices
\begin{eqnarray*}
M^{zz}_\NSR &=& \begin{pmatrix} 
R^{zz}_\NSR & S^{zz}_\NSR \\
-(S^{zz}_\NSR)^\text{T} & R^{zz}_\NSR
\end{pmatrix} ~, \\
N^{zz}_\NSR &=& \begin{pmatrix} 
J^{zz} \!+\! R^{zz}_\NSR & S^{zz}_\NSR \\
(U^{zz}_\NSR)^\text{T} & J^{zz} \!+\! R^{zz}_\NSR
\end{pmatrix} ~,
\end{eqnarray*}
and $4(L - 2\ell) \times 4(L - 2\ell)$ matrix
\begin{eqnarray*}
\Phi^{zz}_\NSR &=& \begin{pmatrix}
M^{zz}_\NSR & N^{zz}_\NSR \\
-(N^{zz}_\NSR)^\text{T} & M^{zz}_\NSR
\end{pmatrix} ~.
\end{eqnarray*}
Finally, we also define $2(L - 2\ell) \times 2L$ matrix
\begin{eqnarray*}
Q^{zz}_\NSR &=&
\begin{pmatrix}
[{\tt AA}]^{m=1\!+\!\frac{L}{2}:L\!-\!\ell}_{n=0:L\!-\!1}(t) & [{\tt AB}]^{m=1\!+\!\frac{L}{2}:L\!-\!\ell}_{n=0:L\!-\!1}(t) \\
[{\tt BA}]^{m=\frac{L}{2}:L\!-\!\ell\!-\!1}_{n=0:L\!-\!1}(t) & [{\tt BB}]^{m=\frac{L}{2}:L\!-\!\ell\!-\!1}_{n=0:L\!-\!1}(t) \\
[{\tt I}]^{m=0:\frac{L}{2}\!-\!\ell\!-\!1}_{n=0:L\!-\!1} & [{\tt AB}]^{m=1:\frac{L}{2}\!-\!\ell}_{n=0:L\!-\!1} \\
[{\tt BA}]^{m=0:\frac{L}{2}\!-\!\ell\!-\!1}_{n=0:L\!-\!1} & -[{\tt I}]^{m=0:\frac{L}{2}\!-\!\ell\!-\!1}_{n=0:L\!-\!1}
\end{pmatrix} ~.
\end{eqnarray*}
We can now compactly write the results of applying the Wick's theorem: 
\begin{eqnarray*}
&& \la 
\sigma^z_{\frac{L}{2}}(t) \sigma^z_{L\!-\!\ell}(t) \sigma^z_0\sigma^z_{\frac{L}{2}\!-\!\ell} 
\sigma^z_{\frac{L}{2}}(t) \sigma^z_{L\!-\!\ell}(t) \sigma^z_0\sigma^z_{\frac{L}{2}\!-\!\ell} 
\ra_\NSR \\
&=& \text{Pf}(\Phi^{zz}_\NSR) ~, \\
&& \la 
\sigma^z_{\frac{L}{2}}(t) \sigma^z_{L\!-\!\ell}(t) \sigma^z_0\sigma^z_{\frac{L}{2}\!-\!\ell}
\sigma^z_{\frac{L}{2}}(t) \sigma^z_{L\!-\!\ell}(t) \sigma^z_0\sigma^z_{\frac{L}{2}\!-\!\ell} 
(-1)^{N_\text{tot}} \ra_\NSR \\
&=& (-1)^{\frac{L(L-1)}{2}} \text{Pf}
\begin{pmatrix}
M^{zz}_\NSR & N^{zz}_\NSR & Q^{zz}_\NSR \\
-(N^{zz}_\NSR)^\text{T} & M^{zz}_\NSR & Q^{zz}_\NSR \\
-(Q^{zz}_\NSR)^\text{T} & -(Q^{zz}_\NSR)^\text{T} & F_\NSR
\end{pmatrix} ~.
\end{eqnarray*}
We evaluate these numerically and combine to obtain the $C_{zz}$ results for the spin system with periodic boundary conditions presented in the main text. 

To calculate $G_{zz}(\ell, t) \equiv \frac{\partial \ln C_{zz}(\ell, t)}{\partial t}$, we use 
\begin{equation*}
G_{zz}(\ell, t) = \frac{\mp 1}{1 \mp \re\sqrt{\text{Pf}[\Phi^{zz}_\text{NS}]}} \re \left(\frac{d\sqrt{\text{Pf}[\Phi^{zz}_\text{NS}]}}{dt} \right) ~,
\end{equation*}
where the upper/lower sign corresponds to the upper/lower sign in $\re F_{zz} = \pm \re \sqrt{\text{Pf}[\Phi^{zz}_\text{NS}]}$ respectively
(recall from the main text that we are calculating $\text{Pf}[\Phi^{zz}_\text{NS}] \approx F_{zz}^2$ and recover the sign when taking the square-root by continuity in parameters $t$ and $\ell$).
We calculate the derivative of the Pfaffian in the standard way, 
\begin{equation*}
\frac{d\text{Pf}[\Phi^{zz}_\text{NS}]}{dt} = \frac{1}{2}\text{Pf}[\Phi^{zz}_\text{NS}] \, \text{Tr}\left[(\Phi^{zz}_\text{NS})^{-1} \frac{d\Phi^{zz}_\text{NS}}{dt} \right] ~.
\end{equation*}
Again, there is essentially no difference between results from $\text{NS}$ sector and from both sectors in the thermodynamic limit.

\section{Pfaffian calculation of $F_{zx}(\ell, t)$} \label{app:Pf_Fzx}
Here we present details of the calculation of $F_{zx}(\ell, t)$. 
The ``doubled'' OTOC $\Gamma_{zx}(\ell, t; L)$, Eq.~(\ref{eqn:Gamma_zx}), can be written in terms of the JW fermions as 
\begin{eqnarray*}
\Gamma_{zx}(\ell, t; L) &=& \left\la \left(\prod_{j=L/2}^{L\!-\!\ell\!-\!1} B_j(t) A_{j\!+\!1}(t) \right) A_0 A_{\frac{L}{2}\!-\!\ell} B_0 B_{\frac{L}{2}\!-\!\ell} \right. \\
&& \left. \left(\prod_{j=L/2}^{L\!-\!\ell\!-\!1} B_j(t) A_{j\!+\!1}(t) \right) A_0 A_{\frac{L}{2}\!-\!\ell} B_0 B_{\frac{L}{2}\!-\!\ell}  \right\ra ~.
\end{eqnarray*}

We define $(L - 2\ell + 4) \times (L - 2\ell + 4)$ matrices 
\begin{widetext}
\begin{eqnarray*}
M^{zx}_\NSR &=& \begin{pmatrix}
[\tt 0] & [{\tt AB}]_{n=0:\frac{L}{2}\!-\!\ell\!-\!1}^{m=1:\frac{L}{2}\!-\!\ell} & [{\tt AA}]^{m=1\!+\!\frac{L}{2}:L\!-\!\ell}_{n=0}(t) & [{\tt AA}]^{m=1+\frac{L}{2}:L\!-\!\ell}_{n=\frac{L}{2}\!-\!\ell}(t) & [{\tt AB}]^{m=1+\frac{L}{2}:L\!-\!\ell}_{n=0}(t) & [{\tt AB}]^{m=1+\frac{L}{2}:L\!-\!\ell}_{n=\frac{L}{2}\!-\!\ell}(t) \\
[{\tt BA}]^{m=0:\frac{L}{2}\!-\!\ell\!-\!1}_{n=1:\frac{L}{2}\!-\!\ell} & [\tt 0] & [{\tt BA}]^{m=\frac{L}{2}:L\!-\!\ell\!-\!1}_{n=0}(t) & [{\tt BA}]^{m=\frac{L}{2}:L\!-\!\ell\!-\!1}_{n=\frac{L}{2}\!-\!\ell}(t) & [{\tt BB}]^{m=\frac{L}{2}:L\!-\!\ell\!-\!1}_{n=0}(t) & [{\tt BB}]^{m=\frac{L}{2}:L\!-\!\ell\!-\!1}_{n=\frac{L}{2}\!-\!\ell}(t) \\
-[{\tt AA}]_{n=1+\frac{L}{2}:L\!-\!\ell}^{m=0}(t) & -[{\tt BA}]_{n=\frac{L}{2}:L\!-\!\ell\!-\!1}^{m=0}(t) & 0 & 0 & [{\tt AB}]^{m=0}_{n=0} & [{\tt AB}]^{m=0}_{n=\frac{L}{2}\!-\!\ell} \\
-[{\tt AA}]_{n=1+\frac{L}{2}:L\!-\!\ell}^{m=\frac{L}{2}\!-\!\ell}(t) & -[{\tt BA}]_{n=\frac{L}{2}:L\!-\!\ell\!-\!1}^{m=\frac{L}{2}\!-\!\ell}(t) & 0 & 0 & [{\tt AB}]^{m=\frac{L}{2}\!-\!\ell}_{n=0} & [{\tt AB}]^{m=0}_{n=0} \\
-[{\tt AB}]_{n=1+\frac{L}{2}:L\!-\!\ell}^{m=0}(t) & -[{\tt BB}]_{n=\frac{L}{2}:L\!-\!\ell\!-\!1}^{m=0}(t) & -[{\tt AB}]^{m=0}_{n=0} & -[{\tt AB}]^{m=\frac{L}{2}\!-\!\ell}_{n=0} & 0 & 0 \\
-[{\tt AB}]_{n=1+\frac{L}{2}:L\!-\!\ell}^{m=\frac{L}{2}\!-\!\ell}(t) & -[{\tt BB}]_{n=\frac{L}{2}:L\!-\!\ell\!-\!1}^{m=\frac{L}{2}\!-\!\ell}(t) & -[{\tt AB}]^{m=0}_{n=\frac{L}{2}\!-\!\ell} & -[{\tt AB}]^{m=0}_{n=0} & 0 & 0
\end{pmatrix} ~, \\
N^{zx}_\NSR &=& \begin{pmatrix}
[\tt I] & 
[{\tt AB}]_{n=0:\frac{L}{2}\!-\!\ell\!-\!1}^{m=1:\frac{L}{2}\!-\!\ell} & 
[{\tt AA}]^{m=1+\frac{L}{2}:L\!-\!\ell}_{n=0}(t) &
[{\tt AA}]^{m=1+\frac{L}{2}:L\!-\!\ell}_{n=\frac{L}{2}\!-\!\ell}(t) & 
[{\tt AB}]^{m=1+\frac{L}{2}:L\!-\!\ell}_{n=0}(t) &
[{\tt AB}]^{m=1+\frac{L}{2}:L\!-\!\ell}_{n=\frac{L}{2}\!-\!\ell}(t) \\
[{\tt BA}]^{m=0:\frac{L}{2}\!-\!\ell\!-\!1}_{n=1:\frac{L}{2}\!-\!\ell} &
-[\tt I] &
[{\tt BA}]^{m=\frac{L}{2}:L\!-\!\ell\!-\!1}_{n=0}(t) & 
[{\tt BA}]^{m=\frac{L}{2}:L\!-\!\ell\!-\!1}_{n=\frac{L}{2}\!-\!\ell}(t) & 
[{\tt BB}]^{m=\frac{L}{2}:L\!-\!\ell\!-\!1}_{n=0}(t) &
[{\tt BB}]^{m=\frac{L}{2}:L\!-\!\ell\!-\!1}_{n=\frac{L}{2}\!-\!\ell}(t) \\
[{\tt AA}]_{n=1+\frac{L}{2}:L\!-\!\ell}^{m=0}(-t) & 
[{\tt AB}]_{n=\frac{L}{2}:L\!-\!\ell\!-\!1}^{m=0}(-t) & 1 & 0 & 
[{\tt AB}]^{m=0}_{n=0} & 
[{\tt AB}]^{m=0}_{n=\frac{L}{2}\!-\!\ell} \\
[{\tt AA}]_{n=1+\frac{L}{2}:L\!-\!\ell}^{m=\frac{L}{2}\!-\!\ell}(-t) & 
[{\tt AB}]_{n=\frac{L}{2}:L\!-\!\ell\!-\!1}^{m=\frac{L}{2}\!-\!\ell}(-t) & 0 & 1 & 
[{\tt AB}]^{m=\frac{L}{2}\!-\!\ell}_{n=0} &
[{\tt AB}]^{m=0}_{n=0} \\
[{\tt BA}]_{n=1+\frac{L}{2}:L\!-\!\ell}^{m=0}(-t) & 
[{\tt BB}]_{n=\frac{L}{2}:L\!-\!\ell\!-\!1}^{m=0}(-t) & 
[{\tt BA}]^{m=0}_{n=0} & 
[{\tt BA}]^{m=\frac{L}{2}\!-\!\ell}_{n=0} & -1 & 0 \\
[{\tt BA}]_{n=1+\frac{L}{2}:L\!-\!\ell}^{m=\frac{L}{2}\!-\!\ell}(-t) & 
[{\tt BB}]_{n=\frac{L}{2}:L\!-\!\ell\!-\!1}^{m=\frac{L}{2}\!-\!\ell}(-t) & 
[{\tt BA}]^{m=0}_{n=\frac{L}{2}\!-\!\ell} & 
[{\tt BA}]^{m=0}_{n=0} & 0 & -1
\end{pmatrix} ~,
\end{eqnarray*}
\end{widetext}
and combine these to form $2(L - 2\ell + 4) \times 2(L - 2\ell + 4)$ matrix 
\begin{eqnarray*}
\Phi_\NSR^{zx} &=& \begin{pmatrix}
M_\NSR^{zx} & N_\NSR^{zx} \\
-(N_\NSR^{zx})^\text{T} & M_\NSR^{zx} \\
\end{pmatrix} ~;
\end{eqnarray*}
finally, we also define $(L - 2\ell + 4) \times 2L$ matrix
\begin{eqnarray*}
Q^{zx}_\NSR &=&
\begin{pmatrix}
[{\tt AA}]^{m=1\!+\!\frac{L}{2}:L\!-\!\ell}_{n=0:L\!-\!1}(t) & [{\tt AB}]^{m=1\!+\!\frac{L}{2}:L\!-\!\ell}_{n=0:L\!-\!1}(t) \\
[{\tt BA}]^{m=\frac{L}{2}:L\!-\!\ell\!-\!1}_{n=0:L\!-\!1}(t) & [{\tt BB}]^{m=\frac{L}{2}:L\!-\!\ell\!-\!1}_{n=0:L\!-\!1}(t) \\
[{\tt I}]^{m=0}_{n=0:L\!-\!1} & [{\tt AB}]^{m=0}_{n=0:L\!-\!1} \\
[{\tt I}]^{m=\frac{L}{2}\!-\!\ell}_{n=0:L\!-\!1} & [{\tt AB}]^{m=\frac{L}{2}\!-\!\ell}_{n=0:L\!-\!1} \\
[{\tt BA}]^{m=0}_{n=0:L\!-\!1} & -[{\tt I}]^{m=0}_{n=0:L\!-\!1} \\
[{\tt BA}]^{m=\frac{L}{2}\!-\!\ell}_{n=0:L\!-\!1} & -[{\tt I}]^{m=\frac{L}{2}\!-\!\ell}_{n=0:L\!-\!1}
\end{pmatrix} ~.
\end{eqnarray*}

We can now write the result of applying the Wick's theorem to the calculation of $\Gamma_{zx}$ as
\begin{eqnarray*}
&& \la \sigma^z_{\frac{L}{2}}(t) \sigma^z_{L\!-\!\ell}(t) \sigma^x_0\sigma^x_{\frac{L}{2}\!-\!\ell} \sigma^z_{\frac{L}{2}}(t) \sigma^z_{L\!-\!\ell}(t) \sigma^x_0\sigma^x_{\frac{L}{2}\!-\!\ell} \ra_\NSR \\ 
&& = \text{Pf}[\Phi^{zx}_\NSR] ~, \\
&& \la \sigma^z_{\frac{L}{2}}(t) \sigma^z_{L\!-\!\ell}(t) \sigma^x_0\sigma^x_{\frac{L}{2}\!-\!\ell} \sigma^z_{\frac{L}{2}}(t) \sigma^z_{L\!-\!\ell}(t) \sigma^x_0\sigma^x_{\frac{L}{2}\!-\!\ell} (-1)^{N_\text{tot}} \ra_\NSR \\
&& = (-1)^{\frac{L(L-1)}{2}} \text{Pf}
\begin{pmatrix}
M^{zx}_\NSR & N^{zx}_\NSR &Q^{zx}_\NSR \\
-(N^{zx}_\NSR)^\text{T} & M^{zx}_\NSR &Q^{zx}_\NSR \\
-(Q^{zx}_\NSR)^\text{T} & -(Q^{zx}_\NSR)^\text{T} & F_\NSR
\end{pmatrix} ~.
\end{eqnarray*}

To calculate $G_{zx}(\ell, t) \equiv \frac{\partial \ln C_{zx}(\ell, t)}{\partial t}$, we use 
\begin{equation*}
G_{zx}(\ell, t) = \frac{\mp 1}{1 \mp \re\sqrt{\text{Pf}[\Phi^{zx}_\text{NS}]}}\re \left(\frac{d\sqrt{\text{Pf}[\Phi^{zx}_\text{NS}]}}{dt} \right) ~,
\end{equation*}
where the upper/lower signs correspond to the upper/lower sign in $\re F_{zx} = \pm \re \sqrt{\text{Pf}[\Phi^{zx}_\text{NS}]}$ respectively (the correct sign is determined using continuity considerations).
The derivative of the Pfaffian can be calculated as $\frac{d\text{Pf}[\Phi^{zx}_\text{NS}]}{dt} = \frac{1}{2} \text{Pf}[\Phi^{zx}_\text{NS}] \text{Tr}[(\Phi^{zx}_\text{NS})^{-1}\frac{d\Phi^{zx}_\text{NS}}{dt}]$.

\section{Exact Heisenberg evolution of $\sigma^x(t)$} \label{app:sigmax_evolution}
Following Ref.~\onlinecite{brandt1976}, we can obtain a compact expression for the Heisenberg evolution of $\sigma_j^x(t)$ under the quantum Ising Hamiltonian, Eq.(\ref{eqn:IsingModel}), at the critical coupling $g = 1$.
With this in hand, we can in fact gain more intuition about the commutator functions $C_{xx}(\ell, t)$ and $C_{zx}(\ell, t)$ at $\beta = 0$ from the operator spreading point of view.
We define Majorana fermions $P_{2j} = (\prod_{j'=-\infty}^{j-1} \sigma_{j'}^x) \sigma_j^z$ and $P_{2j+1} = -(\prod_{j'=-\infty}^{j-1} \sigma_{j'}^x) \sigma_j^y$.
(Note that these are simply related to the Majoranas in the main text and the previous appendices by $A_j = -P_{2j}$ and $B_j = i P_{2j\!+\!1}$;
the convenience of $P_n$'s is that the critical Ising model gives a Majorana chain that is invariant under translation by one Majorana, $n \to n\!+\!1$.)
We have\cite{brandt1976}
\begin{eqnarray}
P_n(t) &=& \sum_k P_{n+k} J_{-k}(2t) = \sum_m P_m J_{n-m}(2t) ~,\\
\sigma_0^x(t) &=& \sum_{m, m'} i P_m P_{m'} J_{-\!m\!+\!1}(2t) J_{-\!m'}(2t) ~,
\label{eqn:sigmax_evol}
\end{eqnarray}
where $J_n$ is the $n$-th order Bessel function of the first kind.

The summation is over all integers $m$ and $m'$ and this expression holds in an infinite system.
We can reorganize the summation over $m$ and $m'$ into a summation over ordered pairs,
\begin{equation}
\sigma_0^x(t) = \sum_{m < m'} i P_m P_{m'} F_{m, m'}(2t) ~,
\end{equation}
where
\begin{equation}
F_{m, m'}(2t) \equiv J_{-\!m\!+\!1}(2t) J_{-\!m'}(2t) - J_{-\!m'\!+\!1}(2t) J_{-\!m}(2t) ~.
\end{equation}
Note that $F_{m, m'}(2t) = -F_{m', m}(2t)$ is antisymmetric.
The summation terms in Eq.~(\ref{eqn:sigmax_evol}) with $m \!=\! m'$ give zero since 
$\sum_m J_{-m\!+\!1} J_{-m} = -\sum_m J_{m\!-\!1} J_m = -\sum_{\tilde{m}} J_{-\tilde{m}} J_{-\tilde{m}\!+\!1} = 0$, where we first used the property $J_{-n} = (-1)^n J_n$ and then changed the summation variable.

Note that the operator $i P_m P_{m'}$ in terms of spin operators is basically a Pauli string of the form $\sigma^{y/z} \sigma^x \sigma^x \dots \sigma^x \sigma^x \sigma^{y/z}$, i.e.,
with $\sigma^x$ in the middle and $\sigma^y$ or $\sigma^z$ at the string ends depending on the parities of $m$ and $m'$; the only exception is $i P_{2j} P_{2j+1} = -\sigma_j^x$.
We can now easily see that the Heisenberg evolution of $\sigma_0^x(t)$ is composed of such Pauli-strings $i P_m P_{m'}$ with amplitudes $F_{m, m'}(2t)$.
This already provides a rough idea of the ``shape'' of the commutator functions $C_{xx}(\ell, t)$ and $C_{zx}(\ell, t)$.
Indeed, since $\sigma_\ell^x$ does not commute with $i P_m P_{m'}$ only when $\ell$ coincides with one of the ends of the string, we expect $C_{xx}$ to have the ``shell-like'' structure\cite{roberts2016} described in the main text.
On the other hand, $\sigma_\ell^z$ does not commute with $i P_m P_{m'}$ when $\ell$ is anywhere inside the string, and this explains the ``dome-like'' structure of $C_{zx}$.

We can supplement these qualitative observations with precise calculations.
The terms in the commutator $[\sigma_0^x(t), \sigma_\ell^x]$ are nonzero when the boundary of the string $i P_m P_{m'}$ hits site $\ell$, which gives us $m = 2\ell$ or $m = 2\ell\!+\!1$ or $m' = 2\ell$ or $m' = 2\ell\!+\!1$, excluding the case $(m = 2\ell, m' = 2\ell\!+\!1)$.
The commutator function $C_{xx}(\ell, t)$ at infinite temperature is easily obtained as the Frobenius norm of $[\sigma_0^x(t), \sigma_\ell^x]$ (divided by $2$).
We therefore have
\begin{eqnarray*}
C_{xx}(\ell, t) &=& 2 \left[ \sum_{m' > 2\ell+1} \!\!\! |F_{2\ell, m'}(2t)|^2 + \!\!\!\! \sum_{m' > 2\ell+1} \!\!\! |F_{2\ell+1, m'}(2t)|^2 \right. \nonumber \\
&& ~~ + \left. \!\! \sum_{m < 2\ell} |F_{m, 2\ell}(2t)|^2 + \!\! \sum_{m < 2\ell} |F_{m, 2\ell+1}(2t)|^2] \right] .
\end{eqnarray*}
With such an expression in hand, we can reproduce the qualitative behavior $C_{xx}(\ell, t) \sim 1/t$ at long times inside the timelike region, $t \gg \ell/c$.
Indeed, it is not difficult to see that 
\begin{equation}
F_{m, m'}(2t) \approx \frac{1}{\pi t} \cos\left[\frac{\pi}{2} (m-m'-1) 
\right] ~, ~\text{for}~|m|, |m'| \ll t ~,
\end{equation} 
while $F_{m, m'}(2t)$ decays quickly once $|m|$ or $|m'|$ exceeds number of order $t$.
This means that the above expression for $C_{xx}(\ell, t)$ contains of order $t$ terms of magnitude of order $1/t^2$, hence $C_{xx}(\ell, t) \sim 1/t$.
A more sophisticated analysis is needed to obtain the amplitude as well as subleading terms, and the treatment in Appendix~\ref{app:Pf_Fxx} provides an alternative derivation giving this data more directly (with the additional benefit of being easily applicable also at finite temperature).
Nevertheless, we find the operator spreading analysis in the present appendix enlightening.

For the commutator function $C_{zx}(\ell, t)$, we can equivalently consider $[\sigma_0^x(t), \sigma_\ell^z]$.
The nonzero contributions come from the $i P_m P_{m'}$ pieces of 
$\sigma_0^x(t)$ with $(m \leq 2\ell, m' \geq 2\ell + 1)$.
This gives us 
\begin{eqnarray}
C_{zx}(\ell, t) &=& 2 \sum_{m \leq 2\ell, m' \geq 2\ell+1} |F_{m, m'}(2t)|^2 ~.
\end{eqnarray}
Using this expression, we can readily understand the finding in the main text that $C_{zx}(\ell, t)$ approaches a nonzero value at long times inside the timelike region, $t \gg \ell/c$.
Indeed, from the behavior of $F_{m, m'}(2t)$ noted earlier, we can see that in the above sum there are of order $t^2$ terms of magnitude of order $1/t^2$, hence nonzero value of the sum in the long-time limit.
Note that the ``operator spreading'' derivation here is much simpler than the formal Pfaffian derivation in Appendix~\ref{app:Pf_Fzx} and gives us almost a closed-form expression for this commutator function at infinite temperature.
On the other hand, the Pfaffian derivation has the advantage of working readily also at finite temperature.

Lastly, we can see different information ``extracted'' from the $\sigma^x(t)$ in other dynamical calculations discussed at the end of Sec.~\ref{subsec:Cxx_longtime}.
For example, the dynamical correlation function at infinite temperature is simply\cite{brandt1976}
\begin{equation}
\langle \sigma_0^x(t) \, \sigma_\ell^x \rangle = -F_{2\ell, 2\ell+1}(2t) \approx \frac{1}{\pi t} ~.  
\end{equation}
We see that the origin of the specific long-time power law behavior in the dynamical correlation function and the OTOC is indeed very different from the operator spreading point of view.

%

\end{document}